\pdfoutput=1 
\documentclass[a4paper, 11pt]{article}
\PassOptionsToPackage{table}{xcolor}

\usepackage{jheppub} % for details on the use of the package, please
\usepackage[T1]{fontenc} % if needed
\usepackage{tangocolors}
\usepackage{tikz}
\usetikzlibrary{arrows, decorations,backgrounds, patterns}
\usetikzlibrary{decorations.pathreplacing ,decorations.markings}
\usepackage{amssymb}
\usetikzlibrary{decorations.pathmorphing,backgrounds,shapes,arrows,shadows}
\usetikzlibrary{patterns.meta}
\usetikzlibrary{patterns,decorations.pathmorphing}
\tikzset{
    snake it/.style={decorate, decoration=snake}
}
\usepackage{pgfplots}
\pgfplotsset{compat=1.11}
\usepgfplotslibrary{fillbetween}
\usetikzlibrary{intersections}
\pgfdeclarelayer{bg}
\pgfsetlayers{bg,main}
\tikzset{zigzag/.style={decorate,decoration=zigzag}}
\tikzset{snake it/.style={decorate, decoration=snake}}
\makeatletter
\def\@hex@@Hex#1%
 {\if a#1A\else \if b#1B\else \if c#1C\else \if d#1D\else
  \if e#1E\else \if f#1F\else #1\fi\fi\fi\fi\fi\fi \@hex@Hex}
\makeatother

\usepackage[all]{xy}
\DeclareFontEncoding{LS2}{}{\noaccents@}
\DeclareFontSubstitution{LS2}{stix}{m}{n}

\DeclareSymbolFont{integrals}{LS2}{stixcal}{m}{n}

\DeclareMathSymbol{\ointupbig}{\mathop}{integrals}{"E8}
\DeclareMathSymbol{\ointupsmall}{\mathop}{integrals}{"B2}

\usepackage[percent]{overpic}
\usepackage{slashed}
\usepackage{wrapfig}
\usepackage{tabu}
\usepackage{diagbox}
\usepackage{mathrsfs,amsmath,amssymb,amsthm,amsfonts,tikz,graphicx,accents,hyperref, color}
\usepackage{dsfont,epiolmec, latexsym, stmaryrd, comment}
\usepackage{slashed,ccaption}
\usepackage{mathrsfs, calligra}
\usepackage{leftidx}
\usepackage{import}
\usepackage{multirow}
\usepackage{amsfonts}
\usepackage{pifont}
\usepackage{tabularx}
\usepackage{cancel}
\usepackage[utf8]{inputenc}
\usetikzlibrary{intersections,calc}
\usepackage{ifthen}
\usepackage{amsmath}
\usepackage{cancel}
\usepackage{caption} 
\usepackage{subcaption}

\usetikzlibrary{patterns.meta}
\usepackage{array}
\DeclareMathOperator{\sech}{sech}

\hypersetup{ linktoc=all,
    colorlinks, linkcolor={palatinateblue},
    citecolor={brightpink}, urlcolor={amaranth}
}

\graphicspath{{Images/}}

\renewcommand{\d}[1]{\ensuremath{\operatorname{d}\!{#1}}}

\def\sideremark#1{\ifvmode\leavevmode\fi\vadjust{\vbox to0pt{\vss% the remark
 \hbox to 0pt{\hskip\hsize\hskip1em%                          will appear only
 \vbox{\hsize2cm\tiny\raggedright\pretolerance10000%          on the side
 \noindent #1\hfill}\hss}\vbox to8pt{\vfil}\vss}}}%
\DeclareSymbolFont{extraup}{U}{zavm}{m}{n}
\DeclareMathSymbol{\varheart}{\mathalpha}{extraup}{86}
\DeclareMathSymbol{\vardiamond}{\mathalpha}{extraup}{87}
\makeatletter
\renewcommand*{\@fnsymbol}[1]{\ensuremath{\ifcase#1\or \clubsuit \or \vardiamond \or \varheart\or
    \spadesuit\or \mathparagraph\or \|\or **\or \dagger\dagger
    \or \ddagger\ddagger \else\@ctrerr\fi}}
\makeatother

\definecolor{rosy}{RGB}{230,235,252}
\definecolor{myframetitle}{RGB}{90,89,170}
\definecolor{myblocktitle}{RGB}{140,185,249}
\definecolor{mytitle}{RGB}{10,80,26}

\definecolor{darkgreen}{RGB}{27,130,45}
\definecolor{darkblue}{rgb}{0,0,0.3}
\definecolor{darkred}{rgb}{0.7,0,0}

\definecolor{light gray}{RGB}{220,220,220}
\definecolor{dark purple}{RGB}{108,0,217}
\definecolor{pink}{RGB}{190,20,100}
\definecolor{orang}{RGB}{193,63,0}
\definecolor{green}{RGB}{11,98,17}
\definecolor{darkpink}{RGB}{153,0,76}
\definecolor{bluegreen}{RGB}{0,102,102}
\definecolor{greenlagan}{RGB}{0,102,0}
\definecolor{redgreen}{RGB}{102,102,0}
\definecolor{Redgreen}{RGB}{153,76,0}
\definecolor{vividviolet}{rgb}{0.62, 0.0, 1.0}
\definecolor{amaranth}{rgb}{0.9, 0.17, 0.31}
\definecolor{palatinateblue}{rgb}{0.15, 0.23, 0.89}
\definecolor{brightpink}{rgb}{1.0, 0.0, 0.5}
\definecolor{cornflowerblue}{rgb}{0.39, 0.58, 0.93}
\definecolor{deepcarminepink}{rgb}{0.94, 0.19, 0.22}
\definecolor{radicalred}{rgb}{1.0, 0.21, 0.37}
\DeclareFontFamily{OT1}{rsfs}{}

\DeclareFontShape{OT1}{rsfs}{m}{n}{ <-7> rsfs5 <7-10> rsfs7 <10->rsfs10}{} 

\DeclareMathAlphabet{\mycal}{OT1}{rsfs}{m}{n}

\newcommand{\be}{\begin{equation}}
\newcommand{\ee}{\end{equation}}
\newcommand{\bea}{\begin{eqnarray}}
\newcommand{\eea}{\end{eqnarray}}
\makeatletter \@addtoreset{equation}{section}

\newtheorem{theorem}{Theorem}
\begin{document}

%%\preprint{IPM/P-2021/nnn}
%\rightline{TUW--21--nn}
%\vskip 1cm
\newcommand{\mytitle}{\begin{center}{\LARGE{Integrability in Three-Dimensional Gravity:}} \\{\Large{\textbf{Eigenfunction-Forced KdV Flows}}}
\end{center}}

\title{{\mytitle}}
\author[a,b]{Hamed~Adami}
\author[c]{,~Anouchah~Latifi}
\affiliation[a]{Center for Mathematics and Interdisciplinary Sciences, Fudan University, Shanghai, 200433, China,}
\affiliation[b]{Shanghai Institute for Mathematics and Interdisciplinary Sciences (SIMIS), Shanghai, 200433, China,}
\affiliation[c]{Department of Mechanics, Qom University of Technology, Iran.}

\emailAdd{hadami@simis.cn}
\emailAdd{latifi@qut.ac.ir}

\abstract{
We uncover a direct connection between three-dimensional gravity with chiral boundary conditions and a class of forced integrable systems. Starting from the Chern--Simons formulation, we derive consistent boundary conditions on a non-compact spatial slice, leading to boundary dynamics described by the potential modified KdV hierarchy. The dynamics reduce to a forced KdV equation, where the forcing term is determined self-consistently by the eigenfunctions of the associated Schr\"{o}dinger operator. Using the inverse scattering transform, the reflectionless sector is solved via the Gelfand--Levitan--Marchenko method, while the radiative sector exhibits universal dispersive decay.
This framework unifies AdS$_3$ boundary dynamics with integrable hierarchies and elucidates the roles of solitons and radiation in the dual conformal field theory.
}

\maketitle

%%%%%%%%%%%%%%%%%%%%%%%%%%%%%%%%%%%%%%%%%%%%%%%%%%%%%%%
\section{Introduction}
%%%%%%%%%%%%%%%%%%%%%%%%%%%%%%%%%%%%%%%%%%%%%%%%%%%%%%%

An \emph{integrable system} is one whose seemingly complicated dynamics can, in fact, be resolved into a collection of independent and predictable motions. In the classical sense, such a system is said to be \emph{solvable by quadratures}, i.e.\ its equations of motion can be expressed in closed form through a finite sequence of algebraic operations and integrations. In contrast to chaotic systems, integrable dynamics exhibit long-term regularity and predictability.

Consider a Hamiltonian system with $N$ degrees of freedom, described on a $2N$-dimensional phase space with canonical coordinates $(q_i,p_i)$. The system is Liouville-integrable if there exist $N$ functionally independent conserved quantities $H_1, H_2, \dots, H_N$
such that $\{ H_i , H_j \} = 0$ for all $i,j$ where $\{\cdot,\cdot\}$ denotes the Poisson bracket. These conserved charges (integrals of motion) constrain the trajectories so that the motion is restricted to an $N$-dimensional invariant manifold in phase space. Under mild regularity assumptions, this manifold is diffeomorphic to an $N$-torus, and the dynamics on it are quasi-periodic. Hence, integrable systems admit exact solutions and exclude chaotic behavior.

The toroidal structure of phase space trajectories provides an intuitive picture. For instance, in two degrees of freedom, the motion can be parametrized by two angular coordinates $(\theta_1,\theta_2)$--the resulting trajectory winds around a two-torus, combining two periodic motions with incommensurate frequencies. Suppose the system is perturbed away from integrability. In that case, this toroidal motion may break down, and the resulting dynamics typically exhibit sensitivity to initial conditions and chaos, as exemplified by the three-body problem.

From a physical perspective, the distinction between integrable and non-integrable dynamics can be summarized as follows:
\begin{itemize}
  \item[(i)] \emph{Solvability:} Integrable models allow exact analytic solutions, whereas generic systems require perturbative or numerical treatment.
  \item[(ii)] \emph{Predictability:} Integrable trajectories are fully predictable, while chaotic dynamics exhibit exponential sensitivity to initial conditions.
  \item[(iii)] \emph{Phase space structure:} Integrable flows are confined to invariant tori, in contrast to the ergodic exploration of higher-dimensional regions typical in non-integrable cases.
  \item[(iv)] \emph{Constants of motion:} Integrable models possess a maximal set of commuting integrals, while generic systems retain only energy and possibly a few additional conserved quantities.
\end{itemize}

The notion of integrability extends beyond finite-dimensional mechanics to field theories and nonlinear partial differential equations. A paradigmatic example is the Korteweg--de Vries (KdV) equation, governing shallow water waves. Such systems exhibit an infinite hierarchy of conserved charges and admit soliton solutions---stable, localized wave packets that preserve their form after scattering. The presence of solitons is a hallmark of integrability in infinite dimensions. Well-known examples include: \emph{1.} The one-dimensional harmonic oscillator, \emph{2.} The Kepler problem (two-body gravitational interaction), \emph{3.} The Calogero--Moser system of interacting particles, \emph{4.} The KdV equation and related soliton--bearing PDEs, \emph{5.} The two-dimensional Ising model at zero external field, and so on.

Integrable systems form rare yet profoundly important examples where exact solvability and mathematical structure coincide. They provide valuable insights into the interplay between symmetry, dynamics, and geometry, and serve as benchmarks against which more general, chaotic systems can be understood.

In recent years, a profound connection has emerged between the study of asymptotic symmetries in three-dimensional gravity and the framework of integrable systems. From the conformal field theory perspective, it has been demonstrated that the BMS$_3$ algebra~\cite{Barnich:2006av,Barnich:2010eb,Bagchi:2014iea,Bagchi:2010zz,Bagchi:2012yk,Bagchi:2013lma,Barnich:2014zoa,Barnich:2014kra,Bagchi:2016bcd,Afshar:2019axx,Yu:2022bcp,Grumiller:2023rzn}—which is isomorphic to the two-dimensional Galilean conformal algebra—can naturally arise within CFT$_2$~\cite{Belavin:1984vu,Cardy:1984bb,Cardy:1986ie} without invoking any explicit flat-space or ultra-relativistic limit. This realization proceeds through a nonlinear construction of the Virasoro generators~\cite{Rodriguez:2021tcz}, whose role is well established in the context of the AdS$_3$/CFT$_2$ correspondence~\cite{Brown:1986nw,Maldacena:1997re,Strominger:1996sh}.

In the gravitational setting, novel families of boundary conditions in AdS$_3$ have been constructed, where the boundary dynamics are governed by integrable hierarchies such as KdV, modified KdV, Gardner, and related systems \cite{compere:2013bya,Perez:2016vqo,Fuentealba:2017omf,Gonzalez:2018jgp,Ojeda:2019xih,Cardenas:2021vwo,Dymarsky:2020tjh,Arenas-Henriquez:2024ypo,Grumiller:2019tyl}. For instance, the choice of $k$-dependent boundary conditions in AdS$_3$ leads to boundary gravitons evolving according to the $k$-th element of the KdV hierarchy, with corresponding anisotropic Lifshitz scaling \cite{Perez:2016vqo}. More generally, suitable boundary conditions allow the Einstein equations in three dimensions to reduce precisely to hierarchies of integrable systems whose symmetries and conserved charges are in one-to-one correspondence with symmetry-generating diffeomorphisms \cite{Fuentealba:2017omf}. 

Dual boundary field theories realizing these dynamics have also been identified, exhibiting non-relativistic features and an infinite set of commuting conserved charges \cite{Gonzalez:2018jgp,Grumiller:2019tyl}. Furthermore, extensions involving the Gardner hierarchy provide a unified framework where both KdV and mKdV structures are simultaneously realized, with direct implications for near-horizon geometries and black hole thermodynamics \cite{Ojeda:2019xih,Grumiller:2019tyl}. Altogether, these results highlight a rich correspondence between three-dimensional gravity, symmetry generators, and the structure of integrable systems.

Integrable methods have proven to be remarkably effective in uncovering nonlinear phenomena in plasma physics, where soliton formation and wave interactions often play a central role. A notable development in this direction is the identification of the caviton \footnote{A caviton is a stable, localized density depletion formed by the nonlinear self-trapping of waves, most famously observed in plasmas.} equations, an integrable class of nonlinear wave equations that capture the dynamics of caviton formation in plasmas \cite{kaup1987integrable, LatifiLeon1991}. These equations admit a solution via the \emph{Inverse Scattering Transform} (IST), yet they display a novel feature compared to more familiar soliton systems: the reflection coefficient is no longer conserved, but may evolve in time, leading either to the amplification of density modulations or to the stabilization of localized soliton-like structures interpreted as cavitons. In parallel, a broad hierarchy of coupled nonlinear evolution equations, including variants such as the coupled KdV--Schr\"{o}dinger system, has been shown to be integrable and solvable under appropriate initial--boundary value conditions \cite{leon1990solution, Latifi2025}. These hierarchies possess B\"{a}cklund transformations and infinitely many conserved quantities, thereby reinforcing the structural link between integrability, conservation laws, and the dynamics of nonlinear excitations in plasma and fluid systems. 

Further insights arise from the study of Langmuir waves coupled to ion acoustic modes through the ponderomotive force, particularly in the Karpman limit, where the dynamics reduce to the so-called caviton equation \cite{kaup1992asymptotic}. Spectral transform techniques demonstrate that long-time evolution generically drives the system toward static reflective configurations or into states supporting permanent localized density depressions, again identifiable as cavitons, comoving with the acoustic background. Taken together, these results underscore a consistent picture: integrable techniques not only ensure mathematical control over nonlinear wave equations but also reveal physically robust structures, such as cavitons, that play a fundamental role in plasma dynamics and nonlinear wave propagation.

One of the central observations of this work is that the deformations introduced above can be naturally understood as arising from different parametrizations of the same underlying phase space. In the Hamiltonian formulation of field theories, the phase space is coordinatized by surface charge densities associated with the asymptotic or boundary symmetries. Their conjugate variables are the corresponding chemical potentials, which enter the variational principle through the specification of boundary conditions and determine which quantities are held fixed at the boundary.

It is important to clearly distinguish two conceptually distinct operations. The first is the choice of boundary conditions, encoded by the selection of chemical potentials as independent control variables. This choice defines the variational problem and, consequently, the physical theory under consideration. The second is a reparametrization of phase space coordinates once a consistent variational principle has been established. Such transformations reorganize the description of the same set of physical configurations without modifying the underlying solution space. From this perspective, a change of slicing should be viewed as a field-space redefinition acting within a fixed phase space structure, rather than as an change of boundary data. Nevertheless, an appropriate parametrization can greatly simplify the implementation and interpretation of given boundary conditions.

In equilibrium settings, it is often convenient to regard chemical potentials as functionals of the surface charge densities. In this case, one may pass between dual descriptions via Legendre transformations, which provide canonical examples of phase space reparametrizations that preserve the physical content while exchanging the roles of conjugate variables.

This viewpoint has a well-established origin in gravitational physics, notably in the work of Kijowski \cite{kijowski1997simple}, where Legendre structures were emphasized in the definition of quasi-local observables. This framework underlies, in particular, the construction of the Liu--Yau quasi-local mass \cite{Liu:2003bx}, which assigns a consistent energy to compact spatial surfaces, see also the review \cite{Szabados:2004xxa}. These developments make explicit that the definition of surface charges and their conjugate potentials is intrinsically tied to the choice of parametrization, while remaining distinct from the specification of boundary conditions themselves.

More recently, these ideas have been further developed in a number of works on gravitational phase space structures and asymptotic dynamics \cite{Grumiller:2019fmp,Adami:2020amw,Adami:2020ugu,Adami:2021sko,Adami:2021nnf,Adami:2022ktn}. These studies clarify how different choices of variables can be implemented consistently within a fixed covariant phase space, thereby refining the separation between boundary data and phase space coordinates. In what follows, we adopt this perspective to reinterpret the deformations introduced above as specific reparametrizations within a unified phase space framework.

The organization of this work is as follows: Section~\ref{sec:2} reviews the first-order Chern--Simons formulation of three-dimensional gravity. We perform the Hamiltonian decomposition, derive the symplectic structure, and present the boundary conditions and associated symmetry algebras that form the foundation of our analysis. Section~\ref{sec:3} establishes the Hamiltonian structure of the boundary theory and demonstrates its integrability. We discuss how a well-posed variational principle leads to a Liouville-integrable system with an infinite hierarchy of mutually commuting conserved charges. Section~\ref{sec:4} introduces a specific restriction of the phase space that connects our general framework to known integrable hierarchies. The recursion relation for the Gelfand--Dikii polynomials is derived, and we show that the boundary dynamics are governed by the KdV hierarchy, with the field $\mathcal{L}_\pm$ serving as the KdV variable. Section~\ref{sec:5} presents the inclusion of a self-consistent spectral forcing. The forcing term $\mathcal{A}_\pm$ is constructed from the eigenfunctions of the associated Schr\"{o}dinger operator $\mathcal{S}_\pm$, resulting in an eigenfunction-forced KdV equation. The integrability of the extended system is established via the construction of modified conserved Hamiltonians. Section~\ref{sec:6} provides explicit solutions of the forced system using the IST. The Gelfand--Levitan--Marchenko (GLM) formalism is employed to reconstruct the potential $\mathcal{L}_\pm$. In this section, we discuss the reflectionless (solitonic) sector, where the one-soliton solution and its holographic interpretation are presented. Section~\ref{sec:7} discuss purely radiative sector, where dispersive long-time asymptotics are analyzed via the stationary phase method. Section~\ref{sec:conc} concludes with a discussion of our results and prospects for future directions, including connections to $T\bar{T}$-like deformations, extensions to flat holography, and the path toward quantization.

%%%%%%%%%%%%%%%%%%%%%%%%%%%%%%%%%%%%%%%%%%%%%%%%%%%%%%%%%%%%%%%%%%%%%%%%%%%%
\section{{First-order formulation of gravity in three dimensions}}\label{sec:2}
%%%%%%%%%%%%%%%%%%%%%%%%%%%%%%%%%%%%%%%%%%%%%%%%%%%%%%%%%%%%%%%%%%%%%%%%%%%%

Three-dimensional gravity can be formulated in the dreibein language by introducing one-forms $e^a$ (dreibein) and $\omega^{ab}$ (spin connection). 
Instead of working solely with the Einstein–Cartan action, we now make explicit use of its Chern--Simons (CS) formulation \cite{Chern:1974ft,Schwarz:1978cn}. It is known that pure gravity with negative cosmological constant in three dimensions can be expressed in terms of two copies of CS theory with two independent $\mathfrak{sl}(2,\mathbb{R})$ Lie-algbera valued connection one-forms, say $A^\pm$ \cite{Achucarro:1986uwr,Witten:1988hc,balasubramanian1999stress,maldacena1999large,Witten:2007kt}.

The Lagrangian $3$-form of the CS theory can be split into a bulk piece plus an exact form
\begin{equation}
	L_{\text{\tiny CS}}[A^\pm] = L_{\text{\tiny bulk}}[A^\pm] + \d L_{\text{\tiny bdy}}[A^\pm] \, , \qquad \text{with} \qquad L_{\text{\tiny CS}}[A^\pm] = \frac{k_\circ}{4\pi} \, \Big\langle A^\pm \wedge \d A^\pm + \frac{2}{3}\, A^\pm \wedge A^\pm \wedge A^\pm \Big\rangle \,,
\end{equation}
where $k_\circ=\ell/4G$ is the CS level and $\langle\cdot,\cdot\rangle$ denotes the invariant bilinear form on $\mathfrak{sl}(2,\mathbb{R})$.
The boundary term $L_{\text{\tiny bdy}}$ does not affect the bulk equations of motion but is crucial for a well-defined variational principle and for the definition of surface charges. In the CS framework, this boundary contribution is related to the appearance of Wess--Zumino--Witten (WZW) terms at the boundary.

Since pure gravity with negative cosmological constant $\Lambda_\circ=-1/\ell^2$ is a topological theory, its action can be rewritten as
\begin{equation}\label{CS-action}
	I_{\text{\tiny EC}}[e,\omega] = I_{\text{\tiny CS}}[A^+] - I_{\text{\tiny CS}}[A^-] \, ,  \qquad \text{with} \qquad I_{\text{\tiny CS}}[A^+] = \int_{\mathcal{M}} L_{\text{\tiny CS}}[A^\pm] \, .
\end{equation}

Let us denote generators of $\mathfrak{sl}(2,\mathbb{R})$ by $L_m$, where $m=\{-1,0,1 \}$. We choose a standard basis to represent these three generators \footnote{Non-vanishing comutators are $\left[L_0,L_{\pm1}\right]=\mp L_{\pm 1}$ and $\left[L_{1},L_{-1}\right]=2L_0$.}
\begin{equation}
    \left[  L_m,L_n\right]= (m-n)\, L_{m+n}\, .
\end{equation}
We also use the invariant bilinear form of $\mathfrak{sl}(2,\mathbb{R})$ in the fundamental representation \footnote{Note that non-vanishing components are $\langle L_1,L_{-1}\rangle=-1$, $\langle L_0,L_0\rangle=1/2$.}
\begin{equation}
    \kappa_{mn}:= \langle L_m L_n\rangle =\begin{pmatrix}
0 & 0 & -1 \\
0 & \frac{1}{2} & 0 \\
-1 & 0 & 0
\end{pmatrix} .
\end{equation}

Taking the first-order variation of the CS Lagrangian we get
\begin{equation}
   \delta L_{\text{\tiny CS}}[A^\pm] = \frac{k_\circ}{2\pi} \, \langle F^\pm \wedge \delta A^\pm \rangle + \d{}\Big\langle - \frac{k_\circ}{4\pi}\,  A^\pm \wedge \delta A^\pm+ \delta L_{\text{\tiny bdy}}[A^\pm] \Big\rangle \, ,
\end{equation}
where $F^\pm= \d A^\pm+A^\pm \wedge A^\pm$ are the field strengths associated with gauge fields $A^\pm$. We can now define symplectic potential components as
\begin{equation}
    \Theta_{\text{\tiny CS}}^\pm = \int_{\mathcal{C}} \Big\langle - \frac{k_\circ}{4\pi}\,  A^\pm \wedge \delta A^\pm+ \delta L_{\text{\tiny bdy}}[A^\pm] +\d Y[\delta A^\pm; A^\pm] \Big\rangle \, ,
\end{equation}
where $Y[\delta A^\pm; A^\pm]$ is a freedom in definition of symplectic potential. 

Since equations of motion, $F^\pm =0$, is covariant under the following gauge transformation $A^\pm \to  (b^\pm)^{-1} (\d +A^\pm ) b^\pm $, we may partially gauge fix to radial gauge and take gauge fields as
\begin{equation}\label{ansatz}
    A^\pm =  (b^\pm)^{-1} \left(\d +a^\pm \right) b^\pm \, ,
\end{equation}
where $b^\pm=b^{\pm}(r)$ are group elements and they depend only on radial coordinate while $a^\pm(t,x)$ depend on coordinates on constant radial slices. Without loss of generality, we may assume that $b^\pm$ are state-independent, that is $\delta b^\pm=0$. Plugging the ansatz \eqref{ansatz} into the symplectic potential, we get
\begin{equation}
    \Theta_{\text{\tiny CS}}^\pm = - \frac{k_\circ}{4\pi}\,\int_{\mathcal{C}} \langle   a^\pm \wedge \delta a^\pm \rangle + \delta \int_{\mathcal{C}} \Big\langle  \frac{k_\circ}{4\pi}\,b^\pm \d (b^\pm)^{-1}\wedge a^\pm+ L_{\text{\tiny bdy}}[A^\pm] \Big\rangle + \int_{\mathcal{C}} \d \,\langle    Y[\delta A^\pm; A^\pm] \rangle \, .
\end{equation}
Note that the first term in the above does not depend on the radial coordinate. Substituting boundary quantities \cite{Grumiller:2016pqb}
\begin{equation}
    a^{\pm}= \left(-\kappa^{mn}\mu_{n}^{\pm} \d t \pm \frac{2\pi}{k_\circ}\,\mathcal{L}^m_\pm \d x \right) L_m \, ,
\end{equation}
into the first term in the symplectic potential, we find
\begin{equation}
    \Theta^\pm = \pm  \int_{\mathcal{C}}   \mu_m^\pm  \, \delta \mathcal{L}^m_\pm \d t \d x \mp \delta \int_{\mathcal{C}} \frac{1}{2}   \mu_m^\pm \, \mathcal{L}^m_\pm\d t \d x \, ,
\end{equation}
and hence symplectic form, which is the exterior derivative of symplectic potential on phase space, reads
\begin{equation}\label{symplectic-form}
    \Omega^\pm = \pm  \int_{\mathcal{C}}  \delta \mu_m^\pm \curlywedge \delta \mathcal{L}^m_\pm\d t \d x \, .
\end{equation}
Equations of motion reduce to 
\begin{equation}\label{EOM-generic}
    \dot{\mathcal{L}}_{\pm}^m \pm \frac{k_\circ}{2\pi}\, \kappa^{mn}\,(\mu_n^\pm)'+\sum_{n \neq -m} \left(m-2\,n \right)\kappa^{np}\mu_{p}^\pm\, \mathcal{L}^{m-n}_\pm  =0
\end{equation}
In global coordinates, AdS$_3$ has a conformal boundary of topology $\mathbb{S}^1 \times \mathbb{R}_t$, with angular coordinate $x\to \ell \phi$, $\phi \sim \phi+2\pi$, and time coordinate $t \in \mathbb{R}$. This corresponds to compact spatial boundary conditions. In contrast, the \emph{Poincaré patch} of AdS$_3$ admits transverse coordinates $t,x \in \mathbb{R}$ and conformal boundary at $r\to\infty$. Here, the boundary spatial section is $\mathbb{R}_x$, so the boundary geometry is non--compact: $\mathbb{R}_x \times \mathbb{R}_t$. We may impose a boundary condition as follows:
\begin{itemize}
  \item decay at the spatial infinity of the boundary (corners at infinity) so that
  \begin{equation}
  \left| \int_{-\infty}^{+\infty} \d x \,\mathcal{L}^m_\pm(t,x) \right| < \infty \, .
\end{equation}
  \item or, one may consider periodic boundary conditions, $x \sim x+L$, under which the mean density $L^{-1} \int_0^L \mathcal{L}_{\pm}^m \, \d x$
can take arbitrary real values.
\end{itemize}
For the latter one, we are primarily interested in finite, non-vanishing values of the mean density, which correspond to nontrivial configurations and include, in particular, the BTZ black hole solutions as special cases.

An infinitesimal version of the gauge transformation  defined above, which preserves the form of $a^\pm$, is given by
\begin{equation}
    \delta_{\epsilon^\pm} a^\pm = \d \epsilon^\pm + \left[ a^\pm ,\epsilon^\pm\right]\, ,
\end{equation}
where $\epsilon^\pm=-\kappa^{mn}\, \epsilon^\pm_n(t,x) L_m $ is the Lie algebra valued parameter of symmetry transformations. Therefore,
\begin{subequations}
    \begin{align}
       & \kappa^{mn}\delta_{\epsilon^\pm} \mu_{n}^{\pm}=\kappa^{mn}\dot{\epsilon}^{\pm}_n +\sum_{n \neq -m}(2n-m)\kappa^{(m-n)p}\kappa^{nq} \mu_p^\pm \epsilon_q^\pm\, ,   \\ &\delta_{\epsilon^\pm} \mathcal{L}^m_\pm=\mp \frac{k_\circ}{2\pi}\, \kappa^{mn} \, (\epsilon^{\pm}_n)' +\sum_{n \neq -m} \left(2\,n-m \right) \kappa^{np}\, \mathcal{L}^{m-n}_\pm \epsilon^{\pm}_p \, .
    \end{align}
\end{subequations}
Also, in the covariant phase space formalism, the surface charge variation associated with the symmetry generator $\epsilon^\pm$ is defined as
\begin{equation}\label{CCV}
    \begin{split}
        \delta Q^\pm(\epsilon^\pm)& = \, \Omega_{\text{\tiny CS}}^\pm [\delta_{\epsilon^\pm} A^\pm , \delta A^\pm] \\
        &= \, \int_{\mathcal{C}} \Big\langle - \frac{k_\circ}{2\pi}\, \delta_{\epsilon^\pm} a^\pm \wedge \delta a^\pm+\d{} \mathcal{Y}[\delta_{\epsilon^\pm} A^\pm,\delta A^\pm; A^\pm] \Big\rangle \\
        &= \, -\frac{k_\circ}{2\pi}\,\int_{\Sigma} \langle  \epsilon^\pm\, \delta a^\pm \rangle +\int_{\Sigma} \Big\langle  \mathcal{Y}[\delta_{\epsilon^\pm} A^\pm,\delta A^\pm; A^\pm] \Big\rangle \, ,
    \end{split}
\end{equation}
where $\Sigma$ denotes a codimension-two surface and the second term is the contribution of the $Y$-term in the surface charge variation,
\begin{equation}
    \mathcal{Y}[\delta_{\epsilon^\pm} A^\pm,\delta A^\pm; A^\pm]= \delta_{\epsilon^\pm} Y[\delta A^\pm; A^\pm]- \delta Y[\delta_{\epsilon^\pm} A^\pm; A^\pm] +Y[\delta_{\delta \epsilon^\pm} A^\pm; A^\pm] \, .
\end{equation}
Since the $Y$-freedom does not affect any of the subsequent analysis, we set it to zero henceforth. The first term in the surface charge variation can be simplified further. To this end, we use explicit expressions for $\epsilon^\pm$ and $a^\pm$. Therefore, we get
\begin{equation}
    \begin{split}
        \delta Q^\pm(\epsilon^\pm)& = \, -\frac{k_\circ}{2\pi}\,\int_{\Sigma} \langle  \epsilon^\pm\, \delta a^\pm \rangle \\
        & =  \pm \int_{\Sigma} \d x\,  \epsilon^\pm_m\, \delta \mathcal{L}_\pm^m \, . 
    \end{split}
\end{equation}
The off-shell Poisson brackets read directly by inverting the symplectic form \eqref{symplectic-form},
\begin{equation}
     \left\{ \mathcal{L}_\pm^m (t,x),\mu^\pm_n(t,y) \right\}_{\text{P}}= \delta^m_n \, \delta(x-y)\,.
\end{equation}
To derive the Dirac bracket, we must utilize the equations of motion.

In constrained Hamiltonian systems, the Dirac bracket is defined entirely in terms of the Poisson brackets of the phase space variables and the second-class constraints. First-class constraints alone cannot be used to define a Dirac bracket, as they generate gauge transformations and lead to a singular constraint matrix. Once a gauge-fixing condition is imposed for each first-class constraint, the resulting pairs become second-class, and the Dirac bracket is then computed with respect to all second-class constraints, including the gauge-fixed first-class constraints. This construction ensures that all constraints can be imposed strongly and that the dynamics automatically respect the chosen gauge (\emph{c.f.} \cite{Brown:1986nw,Carlip:1999cy,Carlip:2008qh}). In the context of three-dimensional gravity, where the theory is topological and has no local dynamical degrees of freedom, the equations of motion themselves correspond to first-class constraints.
Instead, we use the definition of the Dirac bracket in covariant phase space formalism \cite{Lee:1990nz,Iyer:1994ys,Wald:1999wa,Wald:1999vt}
\begin{equation}\label{D-B-Definition}
    \delta_{\gamma} Q(\epsilon):=\left\{ Q(\epsilon), Q(\gamma)\right\}=Q(\alpha)+K[\epsilon,\gamma] \, ,
\end{equation}
where $K[\epsilon,\gamma]$ is central extension term and  $\alpha$ is an skew-symmetric bi-functional of $\epsilon $ and $ \gamma$. Therefore, if we assume that $\epsilon_m^\pm$ are state-independent, we get on-shell commutation relations as
\begin{equation}\label{enveloping-algebra}
    \left\{ \mathcal{L}_\pm^m(t,x) , \mathcal{L}_\pm^n (t,y)\right\} =  \left(m+2n \right) \mathcal{L}_\pm^{m+n}(t,x)\delta(x-y)  \mp \frac{k_\circ}{2\pi} \kappa^{mn} \delta'(x-y)\, .
\end{equation}

%%%%%%%%%%%%%%%%%%%%%%%%%%%%%%%%%%%%%%%%%%%%%%%%%%%%%%%%%%%
\section{Hamiltonian structure and integrability}\label{sec:3}
%%%%%%%%%%%%%%%%%%%%%%%%%%%%%%%%%%%%%%%%%%%%%%%%%%%%%%%%%%%

Requiring a well-posed action principle on the line enforces \emph{vanishing symplectic flux} through spatial infinity,
\begin{equation}\label{Int-cond}
    \int_{\mathcal{C}}\left( \delta\mu^{+}_{m}\wedge\delta \mathcal{L}^{m}_{+}+\delta\mu^{-}_{m}\wedge\delta \mathcal{L}^{m}_{-}\right) \d t \d x=0 \, .
\end{equation}
Under this condition, the chemical potentials $\mu^{\pm}_{m}$ derive from a Hamiltonian functional of $\mathcal{L}^{m}_{-}, \mathcal{L}^{m}_{+}$ and hence phase space is spanned by six functions $\mathcal{L}^{m}_{\pm}$ subjected to equations of motion \eqref{EOM-generic}. The presumed functional can be explicitly written as
\begin{equation}\label{Functional'}
	H[\mathcal{L}^{m}_{+},\mathcal{L}^{m}_{-}] = \int_{\Sigma} \d x\, \mathcal{H}(\cdots, (\mathcal{L}^m_+)^{\text{\tiny(i)}},\cdots,(\mathcal{L}^m_-)^{\text{\tiny(i)}},\cdots )\,.
\end{equation}
Using the boundary condition, $\mu^{\pm}_{m}$ becomes the Euler–Lagrange derivative of the functional $H$, \footnote{Here $f^{(i)}\equiv \partial_x^{\,i}f$ and the Euler–Lagrange operator is $$\frac{\delta}{\delta \mathcal{L}^{m}_{\pm}}=\sum_{i\ge0}(-\partial_x)^{i}\frac{\partial}{\partial(\mathcal{L}^{m}_{\pm})^{(i)}}.$$}
\begin{equation}\label{mu-E-L}
	\mu_m^\pm =\frac{\delta H}{\delta \mathcal{L}^m_\pm}=\sum_{\text{\tiny i}=0}^{\text{\tiny N}} (-1)^{\text{i}}\, \left(\frac{\partial \mathcal{H}}{\partial (\mathcal{L}^m_\pm)^{\text{\tiny(i)}}}\right)^{\text{\tiny(i)}} \, .
\end{equation}
In fact \eqref{Int-cond} is equivalent to integrability of the symplectic potential. Thus, the symplectic potential becomes
\begin{equation}
    \Theta = -\delta   \int_{\mathcal{C}} \left(  \frac{1}{2}\, \frac{\delta H}{\delta \mathcal{L}^m_+} \, \mathcal{L}^m_+ +\frac{1}{2}\, \frac{\delta H}{\delta \mathcal{L}^m_-} \, \mathcal{L}^m_- -\mathcal{H}\right)\d t \d x \, .
\end{equation} By a redefinition of symmetry generators or by a change of coordinates on phase space,  we can write down the surface charge variation in the following form
\begin{equation}
    \delta Q(\epsilon) = \int_{\Sigma} \d x\,  \left( \hat{\epsilon}_m^+ \delta \mathcal{H}^m_+ +\hat{\epsilon}_m^-\, \delta \mathcal{H}^m_-\right) \, ,
\end{equation}
where the new symmetry generators $\hat{\epsilon}_m^\pm$ and $\mathcal{H}^m_\pm$ are new coordinates on the phase space. They are related to the original one as
\begin{equation}
   \epsilon_m^\pm =\sum_{\text{\tiny i}=0}^{\text{\tiny N}} (-1)^{\text{i}}\, \left(\hat{\epsilon}_n^+ \,\frac{\partial \mathcal{H}^n_+}{\partial (\mathcal{L}^m_\pm)^{\text{\tiny(i)}}}+\hat{\epsilon}_n^- \,\frac{\partial \mathcal{H}^n_-}{\partial (\mathcal{L}^m_\pm)^{\text{\tiny(i)}}} \right)^{\text{\tiny(i)}}\, .
\end{equation}
One may choose $\epsilon^\pm_m = \mu^\pm_m$. Then the surface charge variation is manifestly integrable, implying that the total Hamiltonian is a conserved charge, that is, $Q(\epsilon)=H$. Using equations of motion, one would verify that $\dot{H}=0$. So the existence of $H$ implies that it is also conserved.

Utilizing algebra \eqref{enveloping-algebra}, we can define the bracket of two functionals as
\begin{equation}
    \begin{split}
        \left\{ F,G\right\}=& \, (m+2n) \int \d x \left[   \frac{\delta F}{\delta \mathcal{L}_+^m}\, \mathcal{L}_+^{m+n}\,\frac{\delta G}{\delta \mathcal{L}_+^n}+ \frac{\delta F}{\delta \mathcal{L}_-^m}\, \mathcal{L}_-^{m+n} \,\frac{\delta G}{\delta \mathcal{L}_-^n}\right]\\
        &-\frac{k_\circ}{2\pi}\, \kappa^{mn} \int \d x \left[ \frac{\delta F}{\delta \mathcal{L}_+^m}\, \partial_x \frac{\delta G}{\delta \mathcal{L}_+^n}-\frac{\delta F}{\delta \mathcal{L}_-^m} \, \partial_x\frac{\delta G}{\delta \mathcal{L}_-^n}\right]\, .
    \end{split}
\end{equation}

The system is said to be \emph{Liouville integrable} if there exist $6$ smooth Hamiltonian functionals $H^m_\pm= \int_\Sigma \d x\, \mathcal{H}^m_\pm$ satisfying,
\begin{enumerate}
\item[(i)] $\{H^m_\pm,H\}=0$ (constants of motion),
\item[(ii)] $\{H^m_\pm,H^n_\pm\}=\{H^m_+,H^n_-\}=0$ (mutual involution),
\item[(iii)] the differentials $\delta H^m_\pm$ are linearly independent almost everywhere on phase space (functional independence).
\end{enumerate}

%%%%%%%%%%%%%%%%%%%%%%%%%%%%%%%%%%%%%%%%%%%%%%%%%%%%%%%%%%%%%%%%%%
\section{Restricted phase space}\label{sec:4}
%%%%%%%%%%%%%%%%%%%%%%%%%%%%%%%%%%%%%%%%%%%%%%%%%%%%%%%%%%%%%%%%%%

In this section, we focus on the phase space previously analyzed in \cite{Perez:2016vqo}. For completeness and to establish a consistent framework, we briefly review the relevant results of that work. This setup provides a natural starting point for the spectral deformation introduced in the subsequent sections, allowing for a systematic extension of the underlying structure without the need for additional assumptions.

we consider a restricted solution phase space in which we set
\begin{equation}\label{C-01}
     \mathcal{L}_{\pm}^{\mp1}\equiv  -\mathcal{L}_{\pm}\, , \qquad \mathcal{L}_{\pm}^{0}=0\,, \qquad \mathcal{L}_{\pm}^{\pm1}=\frac{k_\circ}{2\pi}\, .
\end{equation}
Plugging the above into the equations of motion results in
\begin{subequations}
    \begin{align}
        & \mu^\pm_{\mp 1}\equiv \mu^\pm \, , \qquad \mu^\pm_0=\frac{1}{2} \,(\mu^\pm)'\, , \qquad \mu^\pm_{\pm 1}= \frac{2\pi}{k_\circ} \mu^\pm\, \mathcal{L}_\pm +\frac{1}{2}\, (\mu^\pm)'' \, ,\\
        & \dot{\mathcal{L}}_\pm = \pm \mathcal{D}_\pm \mu^\pm  \qquad  \text{with} \qquad \mathcal{D}_\pm :=  \mathcal{L}_{\pm}' +2\,  \mathcal{L}_{\pm} \partial_x -\frac{c}{24 \pi} \, \partial_x^3  \, ,
    \end{align}
\end{subequations}
where $c:= 6 k_\circ$ is the Brown-Henneaux central charge. Similarly, the symmetry generator becomes restricted as well,
\begin{subequations}
    \begin{align}
        & \epsilon^\pm_{\mp 1}\equiv  \epsilon^\pm \, , \qquad \epsilon^\pm_0=\frac{1}{2} \,(\epsilon^\pm)'\, , \qquad \epsilon^\pm_{\pm 1}=\frac{2\pi}{k_\circ}\, \epsilon^\pm\, \mathcal{L}_\pm +\frac{1}{2} \, (\epsilon^\pm)'' \, ,\\
        & \delta_{\epsilon^\pm} \mathcal{L}_\pm = \pm\mathcal{D}_\pm \epsilon^\pm \, , \qquad \delta_{\epsilon^\pm} \mu^\pm =\dot{\epsilon}^\pm \pm  \epsilon^\pm (\mu^\pm)'\mp(\epsilon^\pm)' \mu^\pm \, .
    \end{align}
\end{subequations}
One would read total surface charge variation as
\begin{equation}\label{Charg-pm}
    \delta Q(\epsilon) = \int_{\Sigma} \d x\, \left( \epsilon^+\, \delta \mathcal{L}_+ +   \epsilon^-\, \delta \mathcal{L}_- \right) \, .
\end{equation}
From definition \eqref{D-B-Definition}, one can directly deduce the algebra of the surface charges. In particular, the equal-time brackets of $\mathcal{L}_\pm(t,x)$ take the form
\begin{equation}\label{charge-algebra-vir2}
    \{\mathcal{L}_\pm(t, x),\mathcal{L}_\pm(t, y)\}=\pm \mathcal{D}_\pm\delta(x-y) 
    \, 
    .
\end{equation}

Here, we present a systematic overview of the Lax pair formalism, which forms the foundation of the modern theory of integrable systems. The essence of this framework is the reinterpretation of nonlinear evolution equations as isospectral deformations of linear operators. This perspective not only reveals the hidden algebraic structure of integrable models but also establishes a bridge to powerful analytical tools, such as the IST. Our discussion emphasizes the operator-theoretic setting, thereby underscoring the generality of the approach beyond particular equations.

The study of nonlinear partial differential equations (PDEs) with soliton solutions revealed an unexpected feature: despite their apparent complexity, these equations often admit an underlying linear representation. The decisive step in making this structure manifest was the introduction of the \emph{Lax pair} formalism by Peter Lax \cite{lax1968integrals}. Through this construction, nonlinear dynamics can be recast as the compatibility condition of two linear problems, opening a pathway to exact solvability.

Consider a nonlinear evolution equation for a field $\mathcal{L}_\pm(t,x)$ of the form $\dot{\mathcal{L}}_\pm(t,x)=\pm\mathcal{D}_\pm\mu^\pm$ where $\mathcal{D}_\pm$ denotes a nonlinear differential functional of $\mathcal{L}_\pm$. A \emph{Lax pair} for this equation consists of two linear operators $\mathcal{D}_\pm$ and $\mathcal{S}_\pm$, both depending (possibly nonlinearly) on the field $\mathcal{L}_\pm$, such that the dynamics of $\mathcal{L}_\pm$ is equivalent to the \emph{Lax equation} $\dot{\mathcal{S}}_\pm=\pm\left[ \mathcal{D}_\pm,\mathcal{S}_\pm\right]$. Here, $\dot{\mathcal{S}}_\pm$ is defined via its action on test functions.

The operator $\mathcal{S}_\pm$ is associated with a spectral problem $\mathcal{S}_\pm\psi_\pm = k_\pm^2 \psi_\pm$, while $\mathcal{D}_\pm$ encodes the evolution of eigenfunctions. The Lax equation enforces the crucial property that the spectrum of $\mathcal{S}_\pm$ is preserved under time evolution.

\begin{theorem}[Isospectral Flow]
Suppose $\mathcal{S}_\pm(t,x)$ evolves according to a Lax equation. Then the spectrum of $\mathcal{S}_\pm(t,x)$ is invariant in time.
\end{theorem}
This conservation of spectral data immediately yields an infinite tower of constants of motion. For many PDEs of physical interest, this is precisely the structure underlying Liouville integrability. The Lax pair formulation underpins the IST, a nonlinear analogue of Fourier analysis. Thus, highly nonlinear dynamics can be completely resolved by a chain of linear procedures, with the Lax equation ensuring consistency.

Beyond the original case of the Korteweg-de Vries (KdV) equation, the Lax formalism provides a unifying language for a wide spectrum of integrable systems, both continuous and discrete:
\begin{itemize}
    \item \emph{Conserved quantities:} Trace identities guarantee that $\mathrm{Tr}(\mathcal{S}^n)$ are integrals of motion.
    \item \emph{Hierarchy of flows:} Different choices of $\mathcal{D}$ compatible with a fixed $\mathcal{S}$ yield commuting flows, leading to integrable hierarchies such as the KdV or AKNS systems.
    \item \emph{Hamiltonian structures:} The spectral data act as canonical variables, clarifying the action–angle formulation of integrable PDEs.
    \item \emph{Algebraic structures:} The Lax formalism reveals hidden infinite-dimensional symmetries, often linked to loop algebras or Kac–Moody extensions.
\end{itemize}
In a nutshell, the Lax pair formalism exposes the concealed linear backbone of certain nonlinear PDEs. It provides not only a powerful solution method but also a conceptual framework that connects integrability with spectral theory, Hamiltonian mechanics, and representation theory. Its impact extends well beyond soliton equations, influencing developments in quantum field theory, statistical mechanics, and condensed matter physics. The universality of the approach continues to make it an indispensable tool in modern mathematical physics.

Consider a sequence of differential polynomials (local densities) $\mathcal{R}_{\text{I}}[\mathcal{L}_\pm]$ in a function $\mathcal{L}_\pm(t,x)$ defined by the recursion
\begin{equation}\label{eq:recursion}
 \mathcal{R}'_{\text{I}+1}[\mathcal{L}_\pm]
=\mathcal{D}_\pm\mathcal{R}_{\text{I}}[\mathcal{L}_\pm] \, , \qquad \text{with} \qquad \mathcal{D}_\pm :=  \mathcal{L}_\pm' +2\,  \mathcal{L}_\pm \partial_x -\frac{c}{24 \pi} \, \partial_x^3 \, .
\end{equation}
Equation \eqref{eq:recursion} will be our defining recursion for the sequence $\{\mathcal{R}_{\text{I}}\}_{\text{I}\ge0}$.  Below we make precise what we mean by a solution, show existence/uniqueness (up to additive constants), produce the first few densities, and discuss their use in integrable hierarchies and in constructing conserved quantities.

We regard each $\mathcal{R}_{\text{I}}[\mathcal{L}_\pm]$ as a \emph{local differential polynomial} in $\mathcal{L}_\pm$ and its $x$-derivatives (finite polynomial combination of $\mathcal{L}_\pm, \mathcal{L}_\pm', \mathcal{L}_\pm'',\dots$).  Equation \eqref{eq:recursion} determines $ \mathcal{R}'_{\text{I}+1}$ from $\mathcal{R}_{\text{I}}$. Integrating in $x$ yields $\mathcal{R}_{\text{I}+1}$ up to an arbitrary additive constant. So the uniqueness of $\mathcal{R}_{\text{I}+1}$ as a local density requires a choice of the integration constant.  In practice, one may fix the integration constants by imposing a boundary condition that removes the ambiguity.  Under these standard physical boundary conditions the recursion \eqref{eq:recursion} produces a uniquely defined sequence of local densities.

If $\mathcal{R}_0$ is prescribed to be a local differential polynomial, the right–hand side of \eqref{eq:recursion} is manifestly a total $x$-derivative: indeed the operator $\mathcal{D}_\pm$ maps differential polynomials to differential polynomials and satisfies $\mathcal{D}(\text{poly})=\partial_x(\text{poly})$ because every term produced by $\mathcal{D}_\pm$ is of the form $\partial_x(\cdots)$.  Therefore $\mathcal{R}'_{\text{I}+1}$ is a total derivative, and integrating (with the chosen boundary condition) produces $\mathcal{R}_{\text{I}+1}$ as a differential polynomial.  This proves existence of a sequence $\{\mathcal{R}_{\text{I}}\}_{\text{I}\ge 0}$ by induction.

It is convenient for bookkeeping to assign weights $\deg(\partial_x)=1 $ and $ \deg(\mathcal{L}_\pm)=2$ so that $\deg(\mathcal{L}_\pm')=3$, etc.  The operator $\mathcal D_\pm$ is homogeneous of degree $3$ under this grading.  If $\mathcal{R}_{\text{I}}$ is homogeneous of some degree $d_{\text{I}}$, then $ \mathcal{R}'_{\text{I}+1}$ has degree $d_{\text{I}}+3$ and hence $\mathcal{R}_{\text{I}+1}$ has degree $d_{\text{I}}+2$.  Starting from a homogeneous $\mathcal{R}_0$ one obtains a graded tower of densities, which is the usual structure of the Gelfand--Dikii polynomials appearing in KdV-type hierarchies.

A standard and convenient normalization is to choose $\mathcal{R}_0=1$ and hence $\deg(\mathcal{R}_{\text{I}})=2 \,{\text{I}} $. With this choice, the recursion \eqref{eq:recursion} produces the familiar low-order densities.  From $\mathcal{R}'_1[\mathcal{L}_\pm]=\mathcal{D}(\mathcal{R}_0)=\mathcal{L}_\pm'$ we get $\mathcal{R}_1[\mathcal{L}_\pm]=\mathcal{L}_\pm$
where, with the usual zero-mean or decay boundary condition, we set the constant of integration to be zero. Applying the recursion once more, $ \mathcal{R}'_2[\mathcal{L}_\pm]=\mathcal{D}_\pm(\mathcal{L}_\pm)$, and integrating (fixing the additive constant to zero) yields
\begin{equation}\label{eq:R2}
\mathcal{R}_2[\mathcal{L}_\pm]=-\frac{c}{24\pi}\,\mathcal{L}_\pm''+\frac{3}{2}\mathcal{L}_\pm^2\, .
\end{equation}
Similarly, we get
\begin{equation}
    \mathcal{R}_3[\mathcal{L}_\pm] =\left( \frac{c}{24 \pi}\right)^2 \mathcal{L}_\pm^{\text{\tiny(4)}} - \frac{5\,c}{24 \pi}\left[\mathcal{L}_\pm\,\mathcal{L}'' + \frac{1}{2}(\mathcal{L}_\pm')^2 \right] + \frac{5}{2}\, \mathcal{L}_\pm^3 \, ,
\end{equation}
and so on. The pattern above is the familiar sequence of Gelfand--Dikii polynomials up to the chosen scaling proportional to the central charge.

The recursion \eqref{eq:recursion} is precisely the algebraic backbone of the KdV/Gelfand--Dikii hierarchy. One can define the time flow by
\begin{equation}\label{eq:kdvflow}
\dot{\mathcal{L}}_\pm = \pm \mathcal{R}'_{\text{I}+1}[\mathcal{L}_\pm] \, .
\end{equation}
The compatibility of this flow is guaranteed by the structure of the recursion and the fact that the right–hand sides are variational derivatives of commuting Hamiltonians.  The above considerations imply that we shall take the chemical potential as $\mu^\pm=\mathcal{R}_{\text{I}}[\mathcal{L}_\pm]$ and hence Hamiltonians may be obtained by integrating the following equation
\begin{equation}\label{variation-H-I}
   \delta H^\pm_\text{I} = \int_\Sigma \d x \, \mathcal{R}_\text{I}[\mathcal{L}_\pm] \,\delta  \mathcal{L}_\pm \, .
\end{equation}

Consider the anisotropic Lifshitz scaling transformation 
\begin{equation}
     t \longrightarrow \lambda^{-(2 \text{I}+1)} \, t\, ,\qquad x \longrightarrow \lambda^{-1} x\, , \qquad \mathcal{L}_\pm \longrightarrow \lambda^{2}\, \mathcal{L}_\pm \, , \qquad \mathcal{R}_\text{I}[\mathcal{L}_\pm] \longrightarrow \lambda^{2\text{I}} \,\mathcal{R}_\text{I}[\mathcal{L}_\pm]\, ,
\end{equation}
which leaves the KdV and flow equations invariant. Note that under this transformation, derivatives scale as $\partial_x \to \lambda \partial_x$
and for the Hamitonian we have $H^\pm_\text{I} \to \lambda^{2(\text{I}+1)} H^\pm_\text{I}$. Using Euler's homogeneous function theorem, we get
\begin{equation}
    2\, \text{I} \, \mathcal{R}_\text{I}[\mathcal{L}_\pm] = \sum_{\text{i}} (2+\text{i})\,\mathcal{L}_\pm^{\text{\tiny(i)}}\,   \frac{\partial \mathcal{R}_\text{I}[\mathcal{L}_\pm]}{\partial \mathcal{L}_\pm^{\text{\tiny(i)}}}\, , \qquad 2\, (\text{I}+1) \, \mathcal{H}^\pm_\text{I}[\mathcal{L}_\pm] = \sum_{\text{i}} (2+\text{i})\,\mathcal{L}_\pm^{\text{\tiny(i)}}\,   \frac{\partial \mathcal{H}^\pm_\text{I}[\mathcal{L}_\pm]}{\partial \mathcal{L}_\pm^{\text{\tiny(i)}}}\,.
\end{equation}
To construct Hamiltonian functionals from their variations, we employ the \emph{Poincar\'e homotopy formula} in field space. Consider a variational 1-form \eqref{variation-H-I}
where $\mathcal{R}_{\text{I}}[\mathcal{L}_\pm]$ represents the functional derivative of the Hamiltonian. A \emph{straight-line homotopy} is introduced, connecting the trivial configuration $\mathcal{L}_\pm=0$ to the target field $\mathcal{L}_\pm$ as $\mathcal{L}_\pm(t,x;s)=s\, \mathcal{L}_\pm(t,x)$ where $s\in [0,1]$. The Hamiltonian functional is then reconstructed by integrating along this path:
\begin{equation}
   H^\pm_\text{I} = \int_0^1 \d s\int_\Sigma \d x \, \mathcal{R}_\text{I}[s\mathcal{L}_\pm] \,\mathcal{L}_\pm \, .
\end{equation}
This construction extends the standard Poincar\'e lemma to infinite-dimensional field spaces, providing a canonical method to recover Hamiltonians as exact primitives of closed 1-forms. The straight-line homotopy ensures both simplicity and systematicity, allowing one to explicitly express the Hamiltonian in terms of the field configuration and its variational derivative. Consider polynomials that satisfy the following recursion formula
\begin{equation}
    \mathcal{P}^\pm_{\text{I}+1}=-(\mathcal{P}^\pm_{\text{I}})'+\sum_{\text{J}=1}^{\text{I}-1} \mathcal{P}^\pm_{\text{I}-\text{J}}\mathcal{P}^\pm_{\text{J}} \, , 
\end{equation}
with the initial condition $\mathcal{P}^\pm_{\text{1}}=-12\pi\mathcal{L}_\pm/c$. Then Hamiltonians can be explicitly expressed in terms of these polynomials as \cite{novitskii1996motion}
\begin{equation}
    H^\pm_{\text{I}}= 2\left( -\frac{c}{24\pi}\right)^{\text{I}+1}\int_\Sigma \d x\, \mathcal{P}^\pm_{2\text{I}+1} \, .
\end{equation}
This result ensures \eqref{variation-H-I} is integrable. Using the algebraic relations \eqref{charge-algebra-vir2}, we derive the Dirac bracket for arbitrary functionals of $\mathcal{L}_\pm$,
\begin{equation}
   \left\{ F^\pm, G^\pm\right\} = \int_\Sigma \d x   \left( \frac{\delta F^\pm}{\delta \mathcal{L}_\pm}\right)\,  \mathcal{D}_\pm \left( \frac{\delta G^\pm}{\delta \mathcal{L}_\pm}\right),
   \label{dirac-bracket-Lax}
\end{equation}
where $F^\pm$ and $G^\pm$ are functionals of $\mathcal{L}_\pm$. The evolution equation for $\mathcal{L}_\pm$ under the flow generated by a Hamiltonian $H_{\text{I}}^\pm$ follows directly as
\begin{equation}
   \dot{\mathcal{L}}_\pm = \left\{ \mathcal{L}_\pm, H_{\text{I}}^\pm\right\}.
\end{equation}
The recursive structure of the KdV hierarchy can be analyzed directly at the Hamiltonian level. 
By the definition of the Dirac bracket associated with the two compatible Poisson structures, one finds the identity
\begin{equation}
   \left\{ H^\pm_{\text{I}+1},H^\pm_{\text{J}}\right\} \;=\; \left\{ H^\pm_{\text{I}},H^\pm_{\text{J}+1}\right\}, 
\end{equation}
where $H^\pm_{\text{I}}$ denote the Hamiltonians of the hierarchy in the two sectors. Exploiting the skew-symmetry of the Dirac bracket, this recursive relation implies that 
\begin{equation}
   \left\{ H^\pm_{\text{I}},H^\pm_{\text{J}}\right\} = 0 , \qquad \forall\, \text{I},\text{J} \geq 0.
\end{equation}
This implies that the infinite set of conserved charges $\{H^\pm_{\text{I}}\}$ is mutually in involution. Consequently, the flows generated by these charges commute, satisfying the defining criterion for Liouville integrability of the KdV hierarchy. The existence of this infinite-dimensional abelian algebra under the Dirac bracket thus establishes the complete integrability of the system. In a nutshell, the recursion $\mathcal{R}'_{\text{I}+1}[\mathcal{L}_\pm]=\mathcal{D}_\pm \mathcal{R}_{\text{I}}[\mathcal{L}_\pm]$ can be rewritten in Hamiltonian form using the Hamiltonian densities $\mathcal{R}_{\text{I}}[\mathcal{L}_\pm]$ and hence the bi-Hamiltonian structure guarantees involutivity of the Hamiltonians $\{H^\pm_{\text{I}}\}$.

The above result is not an accident but rather a manifestation of a general structural principle. 
According to Magri's theorem \cite{Magri:1977gn,olver1993applications}, the existence of two compatible Hamiltonian structures endows the system with a bi-Hamiltonian formulation. 
This algebraic setting leads naturally to a recursive scheme---often referred to as the Lenard--Magri chain---that generates an infinite sequence of conserved charges. 
In the case of the KdV equation, this mechanism ensures that all Hamiltonians of the hierarchy are well defined, independent, and in mutual involution. 
Consequently, the bi-Hamiltonian structure provides a unified explanation for the integrability of the KdV hierarchy and other related soliton equations.

It is instructive to highlight that the first few conserved charges in this hierarchy carry clear physical significance. Specifically, they correspond, respectively, to the total momentum, the energy, and the next nontrivial conserved quantity of the system. Explicitly, they are given by
\begin{subequations}
    \begin{align}
        & H^\pm_0 = \int_\Sigma \d x \, \mathcal{L}_\pm \, ,\label{H-0}\\
        & H^\pm_1 = \frac{1}{2} \int_\Sigma \d x \, \mathcal{L}_\pm^2 \, , \label{H-1}\\
        & H^\pm_2 = \frac{1}{2} \int_\Sigma \d x \left[ \mathcal{L}_\pm^3 + \frac{c}{24\pi} \left( \mathcal{L}'_\pm \right)^2 \right] \, . \label{H-2}
    \end{align}
\end{subequations}
Here, $\mathcal{L}_\pm$ denotes the dynamical fields of the theory, $\Sigma$ represents the spatial slice, and $c$ is the central charge accounting for quantum or conformal contributions. The derivative term in $H^\pm_2$ reflects the nontrivial structure of higher-order conserved quantities in the integrable hierarchy. These integrals of motion provide a systematic framework to analyze the system's dynamics and its underlying algebraic structure, \emph{c.f.} \cite{dingemans1994water}.

%%%%%%%%%%%%%%%%%%%%%%%%%%%%%%%%%%%%%%%%%%%%%%%%%%%
\paragraph{Chiral transport ($\text{I}=0$).}
%%%%%%%%%%%%%%%%%%%%%%%%%%%%%%%%%%%%%%%%%%%%%%%%%%%
The flow equation together with the choice $\mu^\pm=1$ implies that the dynamical fields and symmetry parameters are chiral functions,
\begin{equation}
   \mathcal{L}_\pm = \mathcal{L}_\pm(t \pm x) \, , \qquad \epsilon^\pm = \epsilon^\pm(t \pm x) \, .
\end{equation}
The corresponding surface charge variation is integrable and reads
\begin{equation}
    Q(\epsilon) = \int_{\Sigma} \d x \, \left( \epsilon^+ \mathcal{L}_+ + \epsilon^- \mathcal{L}_- \right) \, ,
\end{equation}
while the Dirac bracket of the fields defines the surface charge algebra,
\begin{equation}
    \{\mathcal{L}_\pm(t \pm x), \mathcal{L}_\pm(t \pm y)\} = \pm \mathcal{D}_\pm \, \delta(x-y) \, ,
\end{equation}
where $\mathcal{D}_\pm$ encodes the differential operator structure characteristic of the integrable hierarchy.  

%%%%%%%%%%%%%%%%%%%%%%%%%%%%%%%%%%%%%%%%%%%%%%%%%%%%%%%%%%%%%%%%%%%%%%%%%%%%%%%%%%%
\paragraph{Dual boundary theory.}
%%%%%%%%%%%%%%%%%%%%%%%%%%%%%%%%%%%%%%%%%%%%%%%%%%%%%%%%%%%%%%%%%%%%%%%%%%%%%%%%%%%
The dual description of this chiral system is provided by two scalar fields $\varphi_\pm(t, x)$ on the boundary. Their dynamics is generated by the Hamiltonians
\begin{equation}
    H_\pm[\mathcal{L}_\pm] = \int_{\Sigma} \d x \, \mathcal{H}_\pm[\mathcal{L}_\pm] \, ,
\end{equation}
whose Euler-Lagrange derivatives define the associated chemical potentials,
\begin{equation}
    \mu^\pm = \frac{\delta \mathcal{H}_\pm}{\delta \mathcal{L}_\pm} = \mathcal{R}_{\text{I}}[\mathcal{L}_\pm] \, .
\end{equation}
The fields $\varphi_\pm$ are related to the dynamical fields $\mathcal{L}_\pm$ through the Miura transformation,
\begin{equation}\label{eq:miura'}
    \mathcal{L}_\pm = \frac{c}{24\pi} \left[ \frac{1}{2} (\varphi_\pm')^2 + \varphi_\pm'' \right] \, ,
\end{equation}
and transform under the action of the symmetry generators as
\begin{equation}
    \delta_{\epsilon^\pm} \varphi_\pm = (\varphi_\pm)' \, \epsilon^\pm - (\epsilon^\pm)' \, .
\end{equation}
The equations of motion reduce to the chiral evolution equations
\begin{equation}
   \pm \dot{\varphi}_\pm = \varphi_\pm' \, \mu^\pm - (\mu^\pm)' \, .
\end{equation}
In this case, the surface charge variation takes the form
\begin{equation}
   \delta Q(\epsilon^+,\epsilon^-) = \int_{\Sigma} \d x \, \left( \check{\epsilon}^+ \, \delta \varphi_+ + \check{\epsilon}^- \, \delta \varphi_- \right) \, , 
   \qquad \text{with} \qquad 
   \check{\epsilon}^\pm = - \frac{c}{24\pi} \left[ (\varphi_\pm)' \epsilon^\pm - (\epsilon^\pm)' \right]' \, .
\end{equation}
The fields and charge parameters are again chiral functions,
\begin{equation}
    \varphi_\pm = \varphi_\pm(t \pm x) \, , \qquad \check{\epsilon}^\pm = \check{\epsilon}^\pm(t \pm x) \, ,
\end{equation}
and the Dirac brackets of the boundary fields are
\begin{equation}
    \{\varphi_\pm(t \pm x), \varphi_\pm(t \pm y)\} = - \frac{24\pi}{c} \, \vartheta(x-y) \, ,
\end{equation}
where $\vartheta(x-y)$ denotes the Heaviside step function.\footnote{The Heaviside function is defined as
\begin{equation*}
    \vartheta(x) =
\begin{cases}
+1 , & x>0 \\
0 , & x=0 \\
-1 , & x<0
\end{cases} \, .
\end{equation*}}
This boundary description provides a natural realization of the chiral integrable dynamics, linking the conserved charges of the bulk theory to those of the dual scalar fields.

%%%%%%%%%%%%%%%%%%%%%%%%%%%%%%%%%%%%%%%%%%%%%%%%%%%%%%%%%%%
\paragraph{KdV equation ($\text{I}=1$).}
%%%%%%%%%%%%%%%%%%%%%%%%%%%%%%%%%%%%%%%%%%%%%%%%%%%%%%%%%%%
In a similar way to the previous example, the flow equation yields
\begin{equation}\label{eq:KdV-1}
	\dot{\mathcal{L}}_\pm = \pm\left(3\,\mathcal{L}_\pm\, \mathcal{L}'_\pm - \frac{c}{24 \pi}\, \mathcal{L}'''_\pm\right).   
\end{equation}
This is the seminal KdV equation in our conventions. \footnote{In particular, with the rescaling 
{$u=-\frac{2\pi}{k_\circ}\,\mathcal{L}_\pm$ and $\tau=\mp\frac{k_\circ}{4\pi}\,t$}, 
it becomes {$\partial_\tau u = 6\,u\,u' - u'''$}, i.e., the textbook form up to the dispersive sign convention.} Also, symmetry generators shall satisfy the following equation
\begin{equation}\label{eq:KdV-epsilon-1}
     \dot{\epsilon}^\pm  = \pm \left[ 3\,  \mathcal{L}_{\pm} (\epsilon^\pm)' -\frac{c}{24 \pi} \, (\epsilon^\pm)'''\right] \, .
\end{equation}
Equations \eqref{eq:KdV-1} and \eqref{eq:KdV-epsilon-1} coincide for the choice $\epsilon^\pm = \eta^\pm \mathcal{L}_\pm$ where $\eta^\pm$ are state-independent constants.

The KdV equation \eqref{eq:KdV-1}, serves as one of the canonical models of nonlinear wave dynamics. Originally derived in the context of shallow-water hydrodynamics, it has since acquired a universal role in mathematical physics, appearing whenever weak nonlinearity coexists with weak dispersion. The structure of \eqref{eq:KdV-1} embodies this competition: the quadratic advection term $\mathcal{L}_\pm\, \mathcal{L}'_\pm$ acts to steepen gradients and drive wave breaking, whereas the dispersive contribution $ \mathcal{L}_\pm'''$ tends to spread disturbances by introducing a wavenumber-dependent propagation speed. The remarkable property of the KdV equation is that, under suitable conditions, these opposing effects balance in such a way that nontrivial coherent structures emerge (see \cite{dingemans1994water,dinguemans1997water,dingemans1997water} for further discussion).  

A hallmark of this balance is the existence of solitary waves, or solitons. These solutions travel rigidly with constant shape and amplitude-dependent speed. The fact that the velocity grows linearly with amplitude reflects the nonlinear character of the governing dynamics. Beyond their existence as isolated solutions, solitons of the KdV equation exhibit striking particle-like properties: in multi-soliton interactions, individual waves survive collisions unscathed, apart from calculable phase shifts. This resilience is a defining signature of integrability.  

The mathematical richness of the KdV equation extends well beyond its solitary-wave solutions. It was among the first nonlinear partial differential equations recognized as integrable, possessing an infinite sequence of conserved charges and admitting an exact solution via the IST. This discovery not only provided a rigorous framework for understanding soliton interactions but also inaugurated the modern theory of integrable systems. In this framework, the KdV equation serves as a prototype, with connections to algebraic structures such as Lax pairs, hierarchies of commuting flows, and the Gelfand--Dikii polynomials.  

From a physical standpoint, the universality of Eq. \eqref{eq:KdV-1} has ensured its relevance far beyond the shallow-water setting in which it was first derived. Analogues of KdV dynamics arise in plasma physics through ion-acoustic waves, in condensed matter via density waves in nonlinear lattices, and in optics through the propagation of short pulses in dispersive nonlinear media. In each instance, the KdV framework captures the emergence of coherence from the interplay of nonlinearity and dispersion.

The recursion \eqref{eq:recursion} is tightly related to a Lax pair.  Let us define the Schr\"{o}dinger operator as
\begin{equation}\label{Sch-operator}
    \mathcal{S}_\pm:= -\partial_x^2 +\frac{12\pi}{c}\,\mathcal{L}_\pm\, ,
\end{equation}
Gelfand--Dikii theory shows that the coefficients $\mathcal{R}_{\text{I}}$ arise as the (up to normalization) residues or expansion coefficients of the resolvent or of a fractional power of $\mathcal{S}$ (for example $\mathcal{S}^{m+1/2}$) in a formal pseudo-differential expansion.  Equivalently, the $\mathcal{R}_{\text{I}}$ are the coefficients in the asymptotic expansion of the diagonal resolvent $\langle x|(\mathcal{S}-\lambda)^{-1}|x\rangle$ for large spectral parameter $\lambda$.  This spectral viewpoint is the conceptual reason why the densities $\mathcal{R}_{\text{I}}$ generate commuting flows: they correspond to commuting isospectral deformations of $\mathcal{S}$.

A few clarifying remarks to avoid possible confusion are in order:
\begin{enumerate}
  \item The recursion determines $ \mathcal{R}'_{\text{I}+1}$ uniquely from $\mathcal{R}_{\text{I}}$.  The additive constant in $\mathcal{R}_{\text{I}+1}$ can be fixed by any global condition (decay, periodicity with zero mean, or by specifying the Hamiltonian normalization).
  \item Algebraically it is convenient to consider the $\mathcal{R}_{\text{I}}$ as elements of the differential polynomial algebra $C_{\pm}[\mathcal{L}_{\pm},\mathcal{L}_{\pm}',\mathcal{L}_{\pm}'',\cdots]$.  In this algebra, the integration step corresponds to inverting the operator $\partial_x$; the kernel consists precisely of constants.
  \item The normalization $\mathcal{R}_0=1$ is conventional.  Other normalizations (e.g. $\mathcal{R}_{-1}=1$) are used in different presentations; changing the normalization shifts the indexing and rescales the densities.
\end{enumerate}

%%%%%%%%%%%%%%%%%%%%%%%%%%%%%%%%%%%%%%%%%%%%%%%%%%%%%%%%%%%
\section{Including the spectral forcing in the dual action}\label{sec:5}
%%%%%%%%%%%%%%%%%%%%%%%%%%%%%%%%%%%%%%%%%%%%%%%%%%%%%%%%%%%

To reproduce the forced KdV mover, keep the gravitational Poisson operator $\mathcal{D}_\pm$ and modify the chemical potential by \footnote{Under the stated fall–offs $\mathcal{D}_\pm$ is formally skew–adjoint, $\int f\,\mathcal{D}_\pm g\,\d x = - \int g\,\mathcal{D}_\pm f\,\d x$. We define $\mathcal{D}_\pm^{-1}$ as the right inverse determined by the same boundary convention; then $\mathcal{D}_\pm^{-1}$ is also skew–adjoint in the sense $\int f\,\mathcal{D}^{-1}_\pm g\,\d x = - \int g\,\mathcal{D}^{-1}_\pm f\,\d x$
whenever both integrals are well–defined, this follows from definition $\mathcal{D}_\pm\mathcal{D}^{-1}_\pm=\textit{Identity operator}$.}
\begin{equation}
\mu_\pm = \mathcal{R}_{\text{I}}[\mathcal{L}_\pm] + \mathcal{F}_\pm,\qquad
\mathcal{F}_\pm := \mathcal{D}_\pm^{-1}\partial_x \mathcal{A}_\pm\, ,
\end{equation}
so that the flow equation becomes
\begin{equation}\label{Forced-I}
    \dot{\mathcal{L}}_\pm=\pm \left(\mathcal{R}_{\text{I}+1}[\mathcal{L}_\pm] +\mathcal{A}_\pm \right)' \, .
\end{equation}
The forcing term $\mathcal{F}_\pm$ introduced above cannot be chosen arbitrarily. 
Consistency with the variational principle requires that the chemical potential $\mu_\pm$ 
be obtained as the functional derivative of a well-defined Hamiltonian functional. 
In particular, the deformation induced by $\mathcal{F}_\pm$ must itself admit a 
Hamiltonian origin. This requirement imposes a nontrivial integrability condition 
on the allowed form of the forcing.

To ensure this property, we restrict attention to a class of \emph{admissible} forcing 
terms for which there exists a Hamiltonian functional.
Within this class, the deformation corresponds to adding a well-defined boundary 
contribution to the Hamiltonian, and the resulting dynamics remains compatible 
with the canonical structure governed by the Poisson operator $\mathcal{D}_\pm$. With this construction, the forcing can be interpreted as an external source that deforms the integrable hierarchy while preserving its Hamiltonian formulation. 
The quantity $\mathcal{A}_\pm$ thus plays the role of a \emph{forcing component}.

%%%%%%%%%%%%%%%%%%%%%%%%%%%%%%%%%%%%%%%%%%%%%%%%%%%%%%%%%%%%%%%%%	
\subsection{Eigenfunction--forced extension on real line}
%%%%%%%%%%%%%%%%%%%%%%%%%%%%%%%%%%%%%%%%%%%%%%%%%%%%%%%%%%%%%%%%%

We now import the integrable forcing built from products of eigenfunctions of the Schr\"{o}dinger operator associated with \eqref{eq:KdV-1}. Consider the Schr\"{o}dinger--type (generalized Hill--type) equation
\begin{equation}
		\mathcal{S}_{\pm}\psi_\pm= k_{\pm}^2\psi_\pm\, ,
		\label{GHE}
	\end{equation}
where $k_{\pm} \in \mathbb{C}$ is a complex number (as a result of the Lax equation). Not that if $\psi_\pm$ is a solution of \eqref{GHE} then $\dot{\psi}_\pm\mp\mathcal{D}_\pm\psi_\pm$ is also a solution with the same quantum number $k_\pm$. This equation describes the known Hill equation, provided that $\mathcal{L}_\pm(t,x)$ is periodic in $x$ and and the spectral analysis is Floquet–Bloch.
If $\mathcal{L}_\pm(t,x)$ is not periodic and decays at spatial
infinity, the spectral theory shifts then from the Foquet-Bloch analysis to the Scattering Theory in $\mathbb{R}$. It means that one replaces the Bloch spectra with a continuous scattering spectrum (with possible discrete bound states/solitons).
In the periodic case, one gets band--gaps, while in the non-periodic case, the spectrum is continuous with possible bound states (solitons), and the eigenfunctions in the forcing term are scattering states or bound modes, rather than Bloch waves. 

Since \eqref{GHE} is a second-order linear differential equation, its solutions span
a two-dimensional vector space. Let $\chi_\pm$ and $\phi_\pm$ be two linearly independent solutions (fundamental
pair). We may define a $2\times2$ fundamental matrix
\begin{equation}
    \Psi_\pm= 
	\begin{pmatrix}
		\chi_\pm & \phi_\pm \\
		\chi_\pm' & \phi_\pm'
	\end{pmatrix}\,,
\end{equation}
so that we can convert second-order equation \eqref{GHE} to a first order equation
\begin{equation}
    \Psi_\pm'= \Upsilon_\pm\, \Psi \, , \qquad \text{with} \qquad \Upsilon_\pm= \begin{pmatrix}
		0 & 1 \\
		\frac{12\pi}{c}\mathcal{L}_\pm-k_{\pm}^2 & 0
	\end{pmatrix}\,.
\end{equation}
The Wronskian $W_\pm:=\det{\Psi}_\pm$ is constant on the real line, that is $W'=0$, and hence if the Wronskian does not vanish, we can always choose $W=W(t,k_\pm) \neq 0$.
To extract \eqref{Forced-I} with $\text{I}=1$, consider the eigenfunction–forced KdV,
\begin{equation}\label{Forced-general-KdV}
    \dot{\mathcal{L}}_\pm \mp 3\,\mathcal{L}_\pm\, \mathcal{L}'_\pm \pm \frac{c}{24 \pi}\, \mathcal{L}'''_\pm =\pm \partial_x \int_{\mathbb{C}} \d k_{\pm} \wedge \d{} \bar{k}_{\pm}\,\tilde{g}_\pm(t,k)\, \psi_\pm(t,x; k_{\pm})\, \psi_\pm(t,x; -k_{\pm}) \, ,
\end{equation}
where $\tilde{g}_\pm(t,k_{\pm})$ are real, even spectral weights with $\tilde{g}_\pm(t,k_{\pm})= \int_0^t \d \tau\, \tilde{g}_\pm(\tau,k_{\pm})$ and $k_{\pm}= k_{_{R}}^{\pm}+i\, k_{_{I}}^{\pm}$. Equation \eqref{Forced-general-KdV} describes the nonlinear interaction between a high-frequency wavepacket, $\psi_\pm$, and a single, low-frequency wave, $\mathcal{L}_\pm$. This type of coupled-wave equation is of broad interest in theoretical physics. It is closely related to models in plasma physics that describe the coupling of a high-frequency plasma wavepacket to low-frequency acoustic waves. Note that the general method of solution of the system of equations \eqref{GHE} and \eqref{Forced-general-KdV} has been discussed in \cite{leon1990solution}.

Let us now set $\tilde{g}_\pm(t;k_{\pm})= g_\pm(t;k_{\pm}) \, \tilde{\delta}(k_{_{I}}^{\pm})$, where $\tilde{\delta}(k_{_{I}}^{\pm})$ is distribution function defined as
\begin{equation}
    \int_{\mathbb{C}} \d k_{\pm} \wedge \d{} \bar{k}_{\pm}\,f(t,x;k) \tilde{\delta}(k_{_{I}}^{\pm}) = \int_{-\infty}^{+\infty} \d k_{_{R}}^{\pm}\, f(t,x;k_{_{R}}^{\pm}) \, .
\end{equation}
Therefore, we may define the quadratic spectral density as \cite{fokas2022nonlinear,fokas2023korteweg}
\begin{equation}\label{d-h-def}
    \mathcal{A}_\pm :=2\int_{0}^{+\infty}  \psi_{\pm}(t,x;k_{\pm}) \psi_{\pm}(t,x;-k_{\pm}) \, g(t;k_{\pm})\, \d k_{\pm} \, ,
\end{equation}
where $k_\pm \in \mathbb{R}_{\ge0}$. To keep $\mathcal{A}_\pm$ real, we shall impose a condition of the form
\begin{equation}\label{reality-condition}
    \overline{{\psi_{\pm}}(t,x;k_{\pm})}= \psi_{\pm}(t,x;-k_{\pm})\, .
\end{equation}

We may assume that $\mathcal{L}_\pm$ decays fast enough at $x \to \mp \infty$ so that we can drop off the term including it at $x \to \mp\infty$.
Consider a solution of \eqref{GHE} which behaves at spatial (corner) boundaries as
\begin{equation}
    \psi_{\pm}(t,x;k)\sim \begin{cases}
	\mathscr{A}_{\pm}(t;k_\pm)\, e^{ik_\pm x}+ \mathscr{B}_{\pm}(t;k_\pm)\, e^{-ik_\pm x}\, , \hspace{1cm} x\to -\infty\\
	\mathscr{C}_{\pm}(t;k_\pm)\, e^{ik_\pm x}+\mathscr{D}_{\pm}(t;k_\pm)\, e^{-ik_\pm x}\, ,\hspace{1.1cm} x\to +\infty
\end{cases} \, ,
\end{equation}
where $\mathscr{A}_{\pm}, \mathscr{B}_{\pm}, \mathscr{C}_{\pm}$ and $\mathscr{D}_{\pm}$ are the scattering coefficients.
By substituting the asymptotic expansions of $\psi_\pm(t,x;k_\pm)$ and $\psi_\pm(t,x;-k_\pm)$ near $x \to \mp\infty$ into the definition \eqref{d-h-def}, one finds
\begin{subequations}
    \begin{align}
    & \mathcal{A}_{\pm}(t,x)\sim 2\int_{0}^{+\infty}  \left(|\mathscr{A}_{\pm}|^2+|\mathscr{B}_{\pm}|^2+2\, \Re[\mathscr{A}_{\pm}\,  \overline{\mathscr{B}_{\pm}}\, e^{2ik_\pm x}]\right)g_{\pm}\, \d k_\pm \qquad \text{at} \, \, \, x \to -\infty\, , \\
        & \mathcal{A}_{\pm}(t,x)\sim 2\int_{0}^{+\infty}  \left(|\mathscr{C}_{\pm}|^2+|\mathscr{D}_{\pm}|^2+2\, \Re[\mathscr{C}_{\pm}\,  \overline{\mathscr{D}_{\pm}}\, e^{2ik_\pm x}]\right)g_{\pm}\, \d k_\pm \qquad \text{at} \, \, \, x \to +\infty\, .
    \end{align}
\end{subequations}

%%%%%%%%%%%%%%%%%%%%%%%%%%%%%%%%%%%%%%%%%%%%%%%%%%%%%%%%%%%%%
\section{The Gelfand--Levitan--Marchenko integral equation}\label{sec:6}
%%%%%%%%%%%%%%%%%%%%%%%%%%%%%%%%%%%%%%%%%%%%%%%%%%%%%%%%%%%%%

This section revisits the GLM integral equation, a fundamental result in one-dimensional inverse scattering theory. Beyond its classical mathematical formulation, the GLM framework reveals a deep connection between causality, spectral analysis, and the hidden linear structure underlying many nonlinear integrable systems. It provides a rigorous prescription for reconstructing a local potential from its scattering data and plays a central role in the modern theory of integrable nonlinear evolution equations through the IST.

Consider the one-dimensional Schr\"{o}dinger operators \eqref{Sch-operator}, where $\mathcal{L}_\pm$ are assumed to be real, sufficiently smooth, and rapidly decaying as $x \to +\infty$.
The \emph{inverse scattering problem} seeks to determine $\mathcal{L}_\pm$ from the corresponding spectral data of the operators $\mathcal{S}_\pm$.
Unlike the direct problem—solving for the spectrum given a potential—the inverse problem is nonlinear and highly nontrivial.
A complete and elegant solution was formulated independently by Gelfand and Levitan \cite{gel1951determination} and by Marchenko \cite{marchenko1950concerning}, who reformulated the reconstruction procedure as a linear Fredholm integral equation.

\paragraph{Scattering data and the Marchenko kernel.}
For potentials defined on the entire real line, the spectral data associated with $\mathcal{S}_\pm$ consist of:
\begin{itemize}
  \item[\emph{(i)}] the \emph{reflection coefficient} $\mathscr{R}_\pm(t,x,k_\pm)$, defined for $k_\pm \in \mathbb{R}$, characterizing the continuous spectrum;
  \item[\emph{(ii)}] a finite set of discrete eigenvalues $E^\pm_n = -(\alpha^\pm_n)^2$, with $\alpha^\pm_n>0$, corresponding to bound states; and
  \item[\emph{(iii)}] the \emph{norming constants} $\beta^\pm_n > 0$, fixing the normalization of the bound-state wavefunctions.
\end{itemize}
The complete scattering information is thus captured in the data set
\begin{equation}
    \mathfrak{S}_\pm = \{ \mathscr{R}_\pm(t,k_\pm), \, \alpha^\pm_n, \, \beta^\pm_n \}_{n=1}^{N}.
\end{equation}
A central quantity in the GLM approach is the \emph{scattering kernel} $F(t,x)$, defined as
\begin{equation}\label{eq:kernel}
  F_\pm(t,x) = \sum_{n=1}^{N} (\beta^\pm_n)^2 e^{-\alpha^\pm_n x}
  + \frac{1}{2\pi} \int_{-\infty}^{+\infty} \mathscr{R}_\pm(t,k_\pm)\, e^{-ik_\pm x} \, \d k_\pm\, ,
\end{equation}
where $\beta^\pm_n$ might be a function of time. This kernel coherently combines the contributions from both discrete and continuous parts of the spectrum and plays the role of an effective ``memory function'' encoding the scattering response.

\begin{theorem}[GML Equation]
Given the scattering data $\mathfrak{S}_\pm$, the potential $\mathcal{L}_\pm$ is uniquely determined by the scattering kernel $F(t,x)$ defined in \eqref{eq:kernel}. The reconstruction proceeds by solving the linear integral equation
\begin{equation}
  K_\pm(t,x, y) + F_\pm(t,x + y)
  + \int_{x}^{+\infty} K_\pm(t,x, z)\, F_\pm(t,z + y)\, \d z = 0,
  \qquad y \ge x,
  \label{eq:GLM}
\end{equation}
for the Marchenko kernel $K_\pm(t,x, y)$, known as the \emph{GLM kernel}.
The potential is then recovered through the local formula
\begin{equation}\label{eq:potential_recon}
  \mathcal{L}_{\pm}(t,x) = -\frac{c}{6\pi}\,  K_\pm'(t,x, x).
\end{equation}
\end{theorem}
The GLM equation emerges naturally from the analytic and asymptotic properties of the Jost solutions of \eqref{Sch-operator}. 
Let $\psi_\pm(t,x;k)$ be the Jost solution asymptotic to a free plane wave as $x \to +\infty$, \footnote{One needs to use the Leibniz rule for differentiation under the integral with variable limits: \begin{equation*}
\partial_x \int_{a(x)}^{b(x)} f(x,y)\, \d y 
= f\big(x,b(x)\big)\, \partial_x b(x) - f\big(x,a(x)\big)\, \partial_x a(x) + \int_{a(x)}^{b(x)} \partial_x f(x,y)\, \d y\, .
\end{equation*}}
\begin{equation}\label{eq:jost}
  \psi_\pm(t,x; k_\pm) =  e^{-ik_\pm x} + \int_{x}^{+\infty} K_\pm(t,x, y) e^{-ik_\pm y } \, \d y  \, ,
\end{equation}
which is of the form of the \emph{Volterra equation of the second kind}. In fact the Jost solution $\psi(t,x;k_\pm)$ for the spectral problem \eqref{GHE} with the standard normalization at $x\to+\infty$ admits the Volterra representation
\begin{equation}\label{Volterra-representation}
	\psi_\pm(t,x;k_\pm) \;=\; e^{-ik_\pm x} +\frac{6}{c} \int_{-\infty}^{+\infty} G_\pm(x,y;k_\pm)\,\mathcal{L}_\pm(t,y)\,\psi_\pm(t,y;k_\pm)\,\d y \, ,
\end{equation}
where $G_\pm(x,y;k_\pm)$ is Green function of the operator $(\partial_x^2+k^2_\pm)$, that is $(\partial_x^2+k^2_\pm)G_\pm=2\pi\delta(x-y)$.

Reality condition \eqref{reality-condition} implies that $K_\pm(t,x, y)$ is a real function. Taking
\begin{subequations}
    \begin{align}
        & \chi_\pm(t,x,k_\pm):= \psi_\pm(t,x; k_\pm)\, ,\\
        & \phi_\pm(t,x,k_\pm):= \psi_\pm(t,x;- k_\pm)\, ,
    \end{align}
\end{subequations}
with $k_\pm \in \mathbb{R}_{\ge 0}$, as two linearly independent solutions, the Wronskian reads $W(t,k_\pm)= 2ik_\pm$. The function $K_\pm(x, y)$ acts as the kernel of a transformation operator mapping the free solution to the full scattering solution. Requiring that $\psi_\pm(x, k_\pm)$ satisfies the Schr\"{o}dinger equation for all $k_\pm$ yields both \eqref{eq:potential_recon} and the hyperbolic constraint
\begin{equation}
    \left( \partial_x^2 - \partial_y^2 - \frac{12\pi}{c}\mathcal{L}_\pm \right) K_\pm(t,x, y) = 0 \,.
\end{equation}
which is a consequence of the GLM equation \eqref{eq:GLM}. The integral equation \eqref{eq:GLM} follows upon imposing the analytic continuation of scattering amplitudes in the complex $k_\pm$-plane and enforcing the \emph{causality condition} that $ K_\pm(x, y) = 0 $ for $ y < x $. 
This condition ensures that the transformed wavefunction at a given point $ x $ depends only on information from the spatial region to its right, reflecting the unidirectional structure of scattering in one dimension. We shall emphasize that the Lax equation implies that the time evolution of $\psi_\pm(t,x;k)$ is governed by
\begin{equation}\label{psi-t-e}
    \dot{\psi}_\pm(t,x;k)=\pm \mathcal{D}_\pm \psi_\pm(t,x;k)\, .
\end{equation}
The Marchenko equation transcends its origin in linear spectral theory. It constitutes the mathematical backbone of the IST, which linearizes the evolution of many nonlinear integrable partial differential equations.
A prototypical example is the KdV equation.
Within the IST framework, the scattering data $\mathfrak{S}_\pm(t)$ evolve linearly in time, while the Marchenko equation reconstructs the nonlinear field $\mathcal{L}_\pm$ at each instant via the analogues of \eqref{eq:GLM} and \eqref{eq:potential_recon}.
In the special case of \emph{reflectionless potentials} $\mathscr{R}_\pm=0$, the scattering kernel $F_\pm(x)$ becomes a finite sum of exponentials, reducing the integral equation to a tractable system of algebraic relations that yield multi-soliton solutions explicitly.

The GLM formalism exemplifies how nonlinear inverse problems can be recast into linear integral equations with manifestly causal structure. The map
\begin{equation}
    \mathfrak{S}_\pm \longrightarrow F_\pm \longrightarrow K_\pm \longrightarrow \mathcal{L}_\pm \, ,
\end{equation}
demonstrates an elegant chain of linear operations culminating in the local reconstruction formula \eqref{eq:potential_recon}.
This structural simplicity reveals the hidden linearity embedded within certain nonlinear systems—a hallmark of integrability. Although generalizations of the GLM framework exist for higher-dimensional operators and non-Schr\"{o}dinger-type systems, the one-dimensional case remains uniquely transparent and continues to serve as a paradigm for the theory of integrable dynamics and spectral reconstruction.

%%%%%%%%%%%%%%%%%%%%%%%%%%%%%%%%%%%%%%%%%%%%%%%%%%%%%%%%%%%%%
\paragraph{Admissible forcing.}
%%%%%%%%%%%%%%%%%%%%%%%%%%%%%%%%%%%%%%%%%%%%%%%%%%%%%%%%%%%%%

We focus on forcing terms of the self-consistent spectral type defined in \eqref{d-h-def}, where the functions $\psi_\pm$ solve the spectral problem \eqref{GHE}. The normalization of $\psi_\pm$ is fixed once and for all and, in particular, is taken to be independent of variations of the dynamical fields $\mathcal{L}_\pm$. Moreover, both the spectral weights $g_\pm$ and the associated Green functions are assumed to be state-independent. This ensures that all variations originate solely from $\mathcal{L}_\pm$, which is crucial for the integrability of the construction.

Starting from the Volterra representation (6.6), its variation with respect to $\mathcal{L}_\pm$ can be solved for $\delta \psi_\pm$. One finds
\begin{equation}\label{variation-psi}
    \delta\psi_\pm(t,x;k_\pm)= \frac{6}{c} \int_{-\infty}^{+\infty} \mathcal{G}_\pm(t,x,z;k_\pm)\,\psi_\pm(t,z;k_\pm)\,\delta \mathcal{L}_\pm(t,z)\, \d z \, .
\end{equation}
Here $\mathcal{G}_\pm$ denotes the dressed Green function, defined as the solution to \footnote{
By construction, $\mathcal{G}_\pm$ incorporates the full dependence on the background $\mathcal{L}_\pm$. Equivalently, it admits the integral (Dyson) representation
\begin{equation*}
     \mathcal{G}_\pm(t,x,y;k_\pm)= G_\pm(t,x,y;k_\pm)+\frac{6}{c} \int_{-\infty}^{+\infty} \d z\, G_\pm(t,x,z;k_\pm)\, \mathcal{L}_\pm(t,z)\, \mathcal{G}_\pm(t,z,y;k_\pm)\, ,
\end{equation*}
where $G_\pm$ is the Green function associated with the undeformed operator.
}
\begin{equation}
       \left(-\mathcal{S}_\pm+k^2_\pm \right)\mathcal{G}_\pm(t,x,z;k_\pm)=2\pi\delta(x-z)\, .
\end{equation}
Using \eqref{variation-psi}, the variation of \eqref{d-h-def} can be written in the linear form
\begin{equation}
   \delta  \mathcal{A}_\pm (t,x)= \int_{-\infty}^{+\infty} \d y \,  \mathcal{K}_{\pm}(t,x,y) \,\delta \mathcal{L}_\pm(t,y) \, ,
\end{equation}
where the kernel $\mathcal{K}_\pm$ is given by
\begin{equation}
   \mathcal{K}_{\pm}(t,x,y) :=\frac{12}{c} \int_{-\infty}^{+\infty} \d k_{\pm}\,  g_\pm(t;k_{\pm})\, \psi_{\pm}(t,x;-k_{\pm}) \,\psi_\pm(t,y;k_\pm)\, \mathcal{G}_\pm(t,x,y;k_\pm)\, .
\end{equation}

Equation \eqref{Forced-I} then implies that the contribution of the forcing term to the Hamiltonian variation takes the form
\begin{equation}\label{H-forced-variation}
    \delta H^{\text{\tiny (forced)}}_\pm =\int_{-\infty}^{+\infty} \d x\,  \mathcal{A}_\pm (t,x) \,\delta \mathcal{L}_\pm(t,x)\, .
\end{equation}

%%%%%%%%%%%%%%%%%%%%%%%%%%%%%%%%%%%%%%%%%%%%%%%%%%%%
\paragraph{Integrability of the self-consistent source.}
%%%%%%%%%%%%%%%%%%%%%%%%%%%%%%%%%%%%%%%%%%%%%%%%%%%%
A necessary and sufficient condition for the existence of a well-defined Hamiltonian functional $H^{\text{\tiny (forced)}}_\pm$ is the integrability condition $\delta^2 H^{\text{\tiny (forced)}}_\pm=0$. In the present setup, this condition is satisfied provided the kernel $\mathcal{K}_\pm(t,x,y)$ is symmetric under exchange of its spatial arguments,
\begin{equation}
    \mathcal{K}_\pm(t,x,y)=\mathcal{K}_\pm(t,y,x)\, .
\end{equation}
This symmetry follows from the self-adjointness properties of the underlying spectral problem \eqref{GHE}, together with the fact that the dressed Green function inherits the symmetry of the differential operator $-\mathcal{S}_\pm + k_\pm^2$. 

Consequently, the forcing defined through \eqref{d-h-def} is admissible in the sense that it induces an integrable deformation of the Hamiltonian structure. In particular, the variation of the Hamiltonian can be consistently integrated in field space, leading to an additional contribution of the form
\begin{equation}
H^{\text{\tiny (forced)}}_\pm = \int_0^1 \d s\int_\Sigma \d x \, \mathcal{A}_\pm[s\mathcal{L}_\pm] \,\mathcal{L}_\pm \, .
\end{equation}
The above expression provides a homotopy integral representation of the Hamiltonian, which is well-defined precisely because the integrability condition is satisfied. The value of $H^{\text{\tiny (forced)}}_\pm$ is independent of the choice of path in field space, and the linear interpolation adopted here is merely a convenient representative.

Note that, for the $\text{I}$-flow defined in \eqref{Forced-I}, the quantity $\mathcal{A}_\pm$ carries scaling weight $2(\text{I}+1)$. Consistency with the spectral representation then requires fixing the normalization of the eigenfunctions $\psi_\pm$ such that $\psi_\pm \to \lambda^{\text{I}+1} \, \psi_\pm$. This choice ensures that the forcing term is compatible with the scaling properties of the underlying hierarchy.

Assuming that the variation of the total Hamiltonian is entirely induced by the generator of time translation, and using the evolution equation \eqref{Forced-I}, one finds that the total Hamiltonian $H^\pm_{\text{I}}+H_{\text{\tiny (forced)}}^\pm$
is conserved along the flow. This follows directly from substituting the flow equation into the time variation of the Hamiltonian.

\subsection{Reflectionless reduction}
%%%%%%%%%%%%%%%%%%%%%%%%%%%%%%%%%%%%%%%%%%%%%%%%

We now consider the reflectionless sector, characterized by vanishing reflection coefficients $\mathscr{R}_\pm = 0$. In this case, the Marchenko equation \eqref{eq:kernel} reduces to a purely discrete sum, leading to multi-soliton dynamics fully determined by the discrete spectral data $\{\alpha^\pm_n, \beta^\pm_n\}_{n=1}^{N}$ \cite{gardner1967method,segur1980solitons,ablowitz1981solitons,faddeev1987hamiltonian,drazin1989solitons,babelon2003introduction,marchenko2011sturm}. Explicitly, the diagonal Marchenko kernel can be expressed as
\begin{equation}
    K_\pm(t,x,x) = \left[\ln{\det(\mathbf{I} + \mathbf{A}^\pm)} \right]' \, ,
\end{equation}
where $\mathbf{I}$ is the $N\times N$ identity matrix, and $\mathbf{A}^\pm$ is a symmetric matrix with entries
\begin{equation}
    \mathbf{A}^\pm_{nm}(t,x) = \frac{\beta_n^\pm \beta_m^\pm}{\alpha_n^\pm + \alpha_m^\pm}\, e^{-(\alpha_n^\pm + \alpha_m^\pm)x} \, .
\end{equation}
The corresponding finite-rank Marchenko kernel is
\begin{equation}
   K_\pm(t,x,y) = -\mathbf{V}_\pm(t,x)^T \left[ \mathbf{I} + \mathbf{A}^\pm(t,x) \right]^{-1} \mathbf{V}_\pm(t,y) \, , 
   \qquad 
   \mathbf{V}_\pm(t,x) = \left(-\beta^\pm_n e^{-\alpha_n^\pm x}\right)_{N \times 1} .
\end{equation}

%%%%%%%%%%%%%%%%%%%%%%%%%%%%%%%%%%%%%%%%%%%%%%%%%%%
\paragraph{Single-soliton sector.}
%%%%%%%%%%%%%%%%%%%%%%%%%%%%%%%%%%%%%%%%%%%%%%%%%%%
For a single bound state ($N=1$), the Marchenko construction simplifies to a rank-one problem with an explicit solution. Denoting the discrete eigenvalue and associated norming constant by
\begin{equation}
     \alpha^\pm_1 = \alpha_\pm \, , \qquad \beta_1^\pm(t) = \zeta_\pm \, e^{\pm \alpha_\pm v_\pm t} \, ,
\end{equation}
the Marchenko kernel in the reflectionless sector reads
\begin{equation}
K_\pm(t,x,y) = -\frac{2 \alpha_\pm \, e^{\pm 2 \alpha_\pm v_\pm t - \alpha_\pm (x+y) + 2 \alpha_\pm x_\circ^\pm}}{1 + e^{\pm 2 \alpha_\pm v_\pm t - 2 \alpha_\pm x + 2 \alpha_\pm x_\circ^\pm}} \, , 
\qquad 
x^\pm_\circ = \frac{1}{\alpha_\pm} \ln\!\left( \frac{\zeta_\pm}{\sqrt{2 \alpha_\pm}} \right), 
\end{equation}
valid for $y \ge x$, and vanishing otherwise. The reconstructed potential $\mathcal{L}_\pm(t,x)$ is obtained via the standard relation \eqref{eq:potential_recon} is
\begin{equation}\label{eq:soliton-profile}
\mathcal{L}_\pm(t,x) = -\frac{c \, \alpha_\pm^2}{6\pi} \, \sech^2(\alpha_\pm \xi_\pm) \, , 
\qquad 
\xi_\pm(t,x) = x \mp v_\pm t - x_\circ^\pm.
\end{equation}
This solution describes a one-soliton profile of the KdV equation \eqref{eq:KdV-1}, propagating without distortion at velocity $v_\pm = c \, \alpha_\pm^2/(6\pi)$. Its amplitude is proportional to $c\, \alpha_\pm^2$, while the width scales as $1/\alpha_\pm$, reflecting the inverse relation between amplitude and spatial extent characteristic of solitons. The soliton center $x_\circ^\pm$ moves along $x = \pm v_\pm t + x_\circ^\pm$.  

The exponential time dependence in $\beta_1^\pm(t)$ encodes the evolution of the bound-state norming constant under the isospectral KdV flow: the discrete eigenvalue $\alpha_\pm$ remains constant, while the associated Jost coefficient acquires a multiplicative factor $e^{\pm \alpha_\pm v_\pm t}$.  

Equation \eqref{eq:soliton-profile} represents a localized, reflectionless excitation on a vanishing background. Its persistence and robustness under time evolution reflect the integrability of the KdV system: the nonlinear dynamics is fully captured by the time evolution of the scattering data, which for discrete eigenvalues is purely multiplicative. In the general $N$-soliton case, each eigenvalue $\alpha_n^\pm$ generates a solitary component with speed $c (\alpha_n^\pm)^2/(6\pi)$, while the matrix structure of $\mathbf{A}^\pm$ encodes the mutual phase shifts during soliton interactions—a signature of elastic soliton scattering.  

The corresponding Schr\"odinger-type equation \eqref{GHE} with potential \eqref{eq:soliton-profile} admits
\begin{itemize}
    \item \emph{Bound state} ($k_\pm^2 = \alpha_\pm^2$):
    \begin{equation}
        \psi_\pm(t,x;k_\pm) \propto \sech(k_\pm \, \xi_\pm).
    \end{equation}
    \item \emph{Scattering states} ($k_\pm^2 \neq \alpha_\pm^2$, $k_\pm^2>0$):
    \begin{equation}
        \psi_\pm(t,x) \propto e^{i k_\pm \, \xi_\pm} \left[ 1 + \frac{i \alpha_\pm}{k_\pm} \tanh(\alpha_\pm \, \xi_\pm) \right].
    \end{equation}
\end{itemize}

The first three nontrivial conserved charges in the chiral KdV hierarchy are,
\begin{subequations}
\begin{align}
H^\pm_0 &= -\frac{c \alpha_\pm}{3\pi}\,, \\
H^\pm_1 &= \frac{c^2 \alpha_\pm^3}{54\pi^2}\, , \\
H^\pm_2 &= -\frac{7 c^3 \alpha_\pm^5}{1620\pi^3}\, .
\end{align}
\end{subequations}
For the one–soliton configuration \eqref{eq:soliton-profile}, the integral converges due to the exponential fall–off of $\mathcal{L}_\pm$, and the result depends solely on the discrete eigenvalue $\alpha_\pm$, which parametrizes the soliton amplitude and width. The cubic dependence $H_1^\pm \propto \alpha_\pm^3$ reflects the scaling structure of the KdV hierarchy, where each higher Hamiltonian is associated with a higher power of $\alpha_\pm$ and encodes the conserved quantities responsible for integrability.

%%%%%%%%%%%%%%%%%%%%%%%%%%%%%%%%%%%%%%%%%%%%%%%%%%
\paragraph{AdS$_3$/CFT$_2$ interpretation.}
%%%%%%%%%%%%%%%%%%%%%%%%%%%%%%%%%%%%%%%%%%%%%%%%%%
In AdS$_3$ gravity, the chiral functions $\mathcal{L}_\pm(t,x)$ determine the boundary metric data and, equivalently, the holographic energy-momentum tensor via the Brown--Henneaux construction. The one-soliton configuration \eqref{eq:soliton-profile} thus represents a localized, right- or left-moving boundary graviton---a coherent wave packet propagating without dispersion along the conformal boundary.

In the dual CFT$_2$, this corresponds to a coherent excitation of the holomorphic (or anti-holomorphic) component of the stress tensor $T^+(\xi_+)$, whose amplitude and speed are fixed by the discrete eigenvalue $\alpha_\pm$ and the central charge $c$. The reflectionless property implies the absence of a continuous spectrum and identifies $\alpha_\pm$ as a quantized parameter labeling a localized excitation that carries definite KdV charges.

The leading conserved charge $H_1^\pm$ is interpreted holographically as the energy, or equivalently the chiral charge, of the gravitational soliton. More generally, classical configurations of the KdV hierarchy on the AdS$_3$ boundary can be viewed as integrable deformations of the dual CFT$_2$ energy-momentum tensor, generated by a tower of commuting Hamiltonians $H_\text{I}^\pm$. Within this framework, the one-soliton solution provides the simplest realization of an integrable, reflectionless excitation of the boundary stress tensor, capturing the propagation of chiral, dispersionless gravitational modes in three-dimensional gravity.

%%%%%%%%%%%%%%%%%%%%%%%%%%%%%%%%%%%%%%%%%%%%%%%%%%%%%%%%%%%%%%%%%%%%%%%%%%%%%%%%%%%%%%%%%%%%%%%%%%%%%%%%%%%%%%%%%%%%%%%%%%%%%%%%%%%%%%%%%%
\section{Radiative sector}
\label{sec:7}
%%%%%%%%%%%%%%%%%%%%%%%%%%%%%%%%%%%%%%%%%%%%%%%%%%%%%%%%%%%%%%%%%%%%%%%%%%%%%%%%%%%%%%%%%%%%%%%%%%%%%%%%%%%%%%%%%%%%%%%%%%%%%%%%%%%%%%%%%%

We now focus on the case where the discrete spectrum is absent. In this regime, all discrete norming constants vanish, $\beta_n^\pm = 0$, and the scattering kernel introduced in \eqref{eq:kernel} contains only its continuous contribution,
\begin{equation}\label{eq:kernel-continuous}
	F_\pm(t,x) = \frac{1}{2\pi}\int_{-\infty}^{+\infty} \mathscr{R}_\pm(t,k_\pm)\, e^{-i k_\pm x}\, \d  k_\pm .
\end{equation}
The function $\mathscr{R}_\pm(t,k_\pm)$ denotes the time-evolved reflection coefficient corresponding to the Jost solutions normalized along the real axis. In the absence of bound states, the analytic structure of the scattering data simplifies considerably: the reflection coefficient is analytic on the real line with no poles in the upper or lower half-plane, and the entire inverse problem is governed solely by the continuous spectral density encoded in $\mathscr{R}_\pm(t,k_\pm)$.

%%%%%%%%%%%%%%%%%%%%%%%%%%%%%%%%%%%%%%%%%%%%%%%%%%%%%%
\subsection{Marchenko integral equations in the continuous case}
%%%%%%%%%%%%%%%%%%%%%%%%%%%%%%%%%%%%%%%%%%%%%%%%%%%%%%

The GLM equations reduce to Volterra-type integral equations on the half-lines, with the kernel \eqref{eq:kernel-continuous} as their only input. For the right problem, they take the form \eqref{eq:GLM}.
\footnote{An analogous relation holds for the left problem, which obeys
\[
K_\pm(t,x,y) + F_\pm(t,x+y) + \int_{-\infty}^{x} K_\pm(t,x,z) F_\pm(t,z+y)\, \d z = 0, 
\qquad y \le x \, .
\]
\vspace{-0.8cm}
}
The integral operator in each equation is of Volterra-type, which guarantees the existence and uniqueness of the Marchenko kernel $K_\pm(t,x,y)$, provided that the scattering kernel $F_\pm$ satisfies mild regularity and decay conditions.  
The Volterra structure also ensures that the solution can be constructed iteratively, leading to a well-defined reconstruction of the potential from the scattering data.  
The Marchenko kernel $K_\pm(t,x,y)$ admits a convergent Neumann series expansion,
\begin{equation}\label{eq:Neumann}
	\begin{split}
	    K_\pm(t,x,y) =&\, -F_\pm(t,x+y) 
	    + \int_x^{\infty}  \d z \, F_\pm(t,x+z) F_\pm(t,z+y) \\
	    & - \int_x^{\infty} \d z \int_x^{\infty} \d w\,
	    F_\pm(t,x+z) F_\pm(t,z+w) F_\pm(t,w+y)
	    + \cdots \, .
	\end{split}
\end{equation}
Under standard smoothness and decay assumptions on $F_\pm$, this series converges absolutely and provides an explicit iterative reconstruction of $K_\pm$.  
The potential $\mathcal{L}_\pm$ is then obtained from the diagonal value of $K_\pm$ through \eqref{eq:potential_recon}.  
Hence, in the absence of discrete eigenvalues, the inverse scattering problem reduces to a sequence of linear operations determined entirely by the continuous spectral data.

Following the standard spectral transform framework, the inverse problem in the reflection-only sector proceeds through the following steps:
\begin{itemize}
    \item[\emph{1.}]\emph{Direct spectral problem}: Solve the stationary Schr\"{o}dinger equation 
    \begin{equation}
	-\psi'' + \frac{12\pi}{c}\,\mathcal{L}_\pm(0,x)\, \psi = k_\pm^2\, \psi \, ,
    \end{equation}
    to determine the reflection coefficient $\mathscr{R}_\pm(0,k_\pm)$ from the initial potential $\mathcal{L}_\pm(0,x)$.  
    In the purely continuous case, the scattering data consist solely of $\mathscr{R}_\pm$, as no bound-state poles are present.
    
    \item[\emph{2.}]\emph{Time evolution}: 
    The temporal dependence of the scattering data follows from the Lax pair structure. For the KdV hierarchy, the reflection coefficient evolves as
    \begin{equation}
	\mathscr{R}_\pm(t,k_\pm) = \mathscr{R}_\pm(0,k_\pm)\,
	e^{\pm i \omega_\pm t} \, ,
    \end{equation}
    where the specific flow in the integrable hierarchy determines the dispersion relation $\omega_\pm=\omega_\pm(k_\pm)$.
    
    \item[\emph{3.}]\emph{Inverse spectral problem}: Construct the scattering kernel $F_\pm(t,x)$ from the evolved reflection coefficient using \eqref{eq:kernel-continuous}, then solve the GLM equation \eqref{eq:GLM} to obtain Marchenko kernel $K_\pm(t,x,y)$.  
    Finally, reconstruct the potential $\mathcal{L}_\pm(t,x)$ from $K_\pm$ via \eqref{eq:potential_recon}. 
\end{itemize}

%%%%%%%%%%%%%%%%%%%%%%%%%%%%%%%%%%%%%%%%%%%%%%%%%%%%%%%
\subsection{Late-time behavior and physical meaning}
%%%%%%%%%%%%%%%%%%%%%%%%%%%%%%%%%%%%%%%%%%%%%%%%%%%%%%%

We now analyze the asymptotic behavior $t \to +\infty$ of solutions reconstructed from purely continuous spectra—a regime characterized by radiative dispersion and the absence of solitary waves. Our goal is to extract the late-time behavior of $\mathcal{L}_\pm(t,x)$ from the formalism developed above.

The essential feature of the asymptotic analysis is the oscillatory integral
\begin{equation}
F_\pm(t, x) = \frac{1}{2\pi} \int_{-\infty}^{\infty} \mathscr{R}_\pm(0, k_\pm)\, e^{-i \xi_\pm (t,x)}  \d  k_\pm \, , 
\qquad 
\xi_\pm (t,x):=k_\pm x \mp \omega_\pm t \, ,
\label{eq:kernel_evolved}
\end{equation}
whose large-$t$ behavior is determined by the method of stationary phase.  
Dominant contributions arise from points where the phase $\xi_\pm(t,x)$ is stationary,
\begin{equation}\label{eq:stationary_condition}
    \frac{\partial \xi_\pm (t,x)}{\partial k_\pm}=0  
    \qquad \Longrightarrow \qquad 
    \frac{\partial \omega_\pm (k_\pm)}{\partial k_\pm}=\pm \frac{x}{t} \, .
\end{equation}
This condition identifies the mapping between the wavenumber $k_\pm$ and the velocity $v_\pm=x/t$ at which a given mode propagates; $\partial_{k_\pm}\omega_\pm$ is the group velocity. Hence, the asymptotic profile is analyzed along the characteristic rays $x=\pm v_\pm t$.

Let $k^\ast_\pm$ denote a stationary point satisfying \eqref{eq:stationary_condition}.  
Assuming $\mathscr{R}_\pm(0,k_\pm)$ is sufficiently smooth and rapidly decaying, the leading-order late-time behavior of \eqref{eq:kernel_evolved} is
\begin{equation}\label{eq:stationary_phase_result}
F_\pm(t, x) \sim 
\frac{\mathscr{R}_\pm(0, k_\pm^\ast)}{\sqrt{2\pi t |\sigma^\ast_\pm|}} 
\exp\!\left[-i \xi^\ast_\pm + i \frac{\pi}{4} \operatorname{sgn}(\sigma^\ast_\pm)\right] 
+\text{c.c.}+ \mathcal{O}(t^{-1}) , 
\qquad 
\sigma^\ast_\pm = 
\frac{\partial^2 \omega_\pm (k_\pm)}{\partial k_\pm^2}\Big|_{k_\pm^\ast},
\end{equation}
where the phase shift $i(\pi/4)\operatorname{sgn}(\sigma^\ast_\pm)$ arises from the Fresnel integral.\footnote{%
The stationary-phase approximation gives
\[
F_\pm(t, x) \sim \frac{1}{2\pi} \mathscr{R}_\pm(0, k_\pm^\ast)\, e^{-i \xi_\pm^\ast} 
\int_{-\infty}^{\infty} 
\exp\!\left[\frac{i}{2} t\,\sigma^\ast_\pm (k_\pm - k_\pm^\ast)^2\right] \d k_\pm,
\]
and the Fresnel integral 
$\int_{-\infty}^{+\infty} e^{i\alpha \xi^2} \d \xi 
= \sqrt{\pi/|\alpha|}\, e^{i(\pi/4)\operatorname{sgn}(\alpha)}$.} 
If several isolated stationary points exist, their contributions must be summed.

At leading order, we have $K_\pm(t,x,x) \sim -F_\pm(t,2x)$, and hence the potential exhibits the universal large-$t$ behavior
\begin{equation}\label{eq:potential_explicit}
\mathcal{L}_{\pm}(t,x) \sim 
-\frac{i\,c\, k_\pm^\ast\, \mathscr{R}_\pm(0, k_\pm^\ast)}{3\pi \sqrt{2\pi t |\sigma^\ast_\pm|}} 
\exp\!\left[-2i k_\pm^\ast x \pm i \omega_\pm^\ast t + i \frac{\pi}{4} \operatorname{sgn}(\sigma^\ast_\pm)\right]
+\text{c.c.}+ \mathcal{O}(t^{-1}),
\end{equation}
where $k_\pm^\ast$ satisfies the group-velocity condition \eqref{eq:stationary_condition}.

The evolution of the potential is governed by the source-free KdV equation.  
Substituting the late-time form \eqref{eq:potential_explicit} into that equation determines the dispersion relation,
\begin{equation}\label{eq:dispersion_relation}
\omega_\pm(k_\pm) = \frac{c\, k_\pm^3}{3\pi} \, .
\end{equation}
For the KdV flow, the group velocity and curvature factor are $v_\pm=c\, k_\pm^2/\pi$ and $\sigma_\pm=2c\, k_\pm/\pi$, respectively.

The asymptotic analysis reveals several universal features:
\begin{itemize}
    \item[\emph{i})] \emph{Decay laws and scaling:}  
    The potential exhibits the characteristic dispersive decay
    \[
    F_\pm(t,\xi) \sim \mathcal{O}(t^{-1/2}), 
    \qquad 
    \mathcal{L}_\pm(t,x) \sim \mathcal{O}(t^{-1/2}) ,
    \]
    which is universal for one-dimensional dispersive systems.

    \item[\emph{ii})] \emph{Group velocity and causality:}  
    The stationary condition \eqref{eq:stationary_condition} defines the causal propagation structure through the group velocity $v_\pm$, which maps wavenumbers to asymptotic spatial regions.

    \item[\emph{iii})] \emph{Asymptotic linearization:}  
    The nonlinear term $\mathcal{L}_\pm \mathcal{L}'_\pm$ becomes subdominant at large $t$, showing that integrable systems linearize asymptotically in the radiative regime—nonlinearity affects only subleading corrections.

    \item[\emph{iv})] \emph{Phase structure:}  
    The universal phase factor $\exp[i\pi/4\,\operatorname{sgn}(\sigma^\ast_\pm)]$ originates from the Fresnel integral and encodes a topological property of the stationary-phase approximation.
\end{itemize}

The analysis above provides a comprehensive description of the late-time dynamics of integrable systems with purely continuous spectra.  
The cubic dispersion relation \eqref{eq:dispersion_relation} characterizes the KdV-type hierarchy, while the $t^{-1/2}$ decay sets the universal relaxation rate for one-dimensional radiative dynamics.  
The stationary-phase framework shows explicitly how the initial reflection data $\mathscr{R}_\pm(0,k_\pm)$ determine the spatial structure of the late-time wave field via the group-velocity mapping.

Physically, this sector corresponds to configurations where the potential does not support localized bound states.  
The field reconstructed through the Marchenko equation is therefore purely radiative and soliton-free.  
Despite the absence of localized excitations, the inverse scattering formalism remains exact, providing a one-to-one mapping between the continuous scattering data and the evolving field.  
In this regime, the conserved quantities of the integrable hierarchy retain their time-independence: each charge is expressed as a spectral integral over the reflection coefficient, whose time evolution merely introduces phase factors that cancel in the modulus.  
Thus, while energy and other conserved densities may be dynamically redistributed across wavenumbers, their integrated values remain invariant under time evolution.

The reflection-only regime also exhibits subtle physical effects.  
The absence of discrete eigenvalues can produce partial or total transparency for certain spectral bands of $\mathscr{R}_\pm$, effectively acting as a spectral filter. Such mechanisms underlie several physical phenomena, including plasma wavepacket dispersion and capillary–gravity wave propagation. \footnote{ Capillary–gravity waves are small-amplitude surface waves at a fluid interface for which both gravity and surface tension act as restoring forces. Their linear dispersion relation
\begin{equation}
    \omega^2 = g\,k + \frac{\sigma}{\rho}\,k^3 ,
\end{equation}
relates the angular frequency $\omega$ to the wavenumber $k$; here $g$ is the gravitational acceleration, $\sigma$ is the surface tension coefficient, and $\rho$ is the fluid density. These waves are dispersive: both the phase velocity $v_p=\omega/k$ and the group velocity $v_g=\partial_k\omega$ depend nontrivially on $k$, with gravity dominating at small $k$ and surface tension at large $k$.

}

When a stationary point approaches $k_\pm=0$ (e.g. along the critical ray $x/t\to 0^+$ for KdV), one has $\sigma_\pm^\ast\to 0$, and the non-degenerate stationary-phase formula \eqref{eq:stationary_phase_result} ceases to apply.  
In this case, a rescaling reduces the phase to an Airy-type integral, yielding the slower decay
\begin{equation}
	F_\pm(t,2x)=\mathcal O(t^{-1/3})\,,\qquad
	\mathcal L_\pm(t,x)=\mathcal O(t^{-1/3})\, ,
\end{equation}
valid uniformly within a $t^{-2/3}$-wide neighborhood of the turning ray.  
Away from this narrow region, the standard $t^{-1/2}$ law holds.

Higher-order terms in the Neumann series \eqref{eq:Neumann} yield convolutions of $F_\pm$ that are subleading in $t$ for short-range data, while derivatives of $k_\ast(x/t)$ contribute only at order $t^{-3/2}$.  
Therefore, \eqref{eq:potential_explicit} represents the complete leading-order asymptotic behavior in the reflection-only sector.

The GLM equation \eqref{eq:GLM}, together with the reconstruction formula \eqref{eq:potential_recon}, provides the \emph{exact nonlinear} inverse map from scattering data to the field.  
The nonlinearity is encoded in the Volterra integral structure and in the convergent Neumann series \eqref{eq:Neumann} that determines $K_\pm$ from $F_\pm$.

The stationary-phase analysis applied to the oscillatory integrals \eqref{eq:kernel-continuous} and \eqref{eq:potential_explicit} captures the large-$t$ limit of the exact reconstruction.  
In the reflection-only case (no discrete spectrum), the leading term of the exact solution becomes linear in the reflection coefficient and obeys the dispersive decay $\mathcal L_\pm(t,x)=\mathcal O(t^{-1/2})$ along generic rays.  
Nonlinear corrections arise only at subleading orders (from higher terms in \eqref{eq:Neumann} and curvature corrections), and are suppressed for short-range data.  
This behavior is known as \emph{asymptotic linearization}: the leading radiation is linear, but it originates from the exact nonlinear inverse-scattering solution.

When a discrete spectrum is present, the full Marchenko reconstruction decomposes as
\begin{equation}
\mathcal L_\pm(x,t) = \mathcal L_\pm^{\text{sol}}(t,x) + \mathcal L_\pm^{\text{rad}}(t,x),
\end{equation}
where $\mathcal L_\pm^{\text{sol}}$ contributes $\mathcal O(1)$ along soliton rays (non-dispersive), while $\mathcal L_\pm^{\text{rad}}$ follows the radiative asymptotics derived above.  
In the purely continuous case discussed in this section, $\mathcal L_\pm^{\text{sol}}(t,x)\equiv 0$, and the radiation asymptotics fully determine the late-time dynamics.

%%%%%%%%%%%%%%%%%%%%%%%%%%%%%%%%%%%%%%%%%%%%%%
\section{Conclusion and outlook}\label{sec:conc}
%%%%%%%%%%%%%%%%%%%%%%%%%%%%%%%%%%%%%%%%%%%%%%

\paragraph{Relation to $T\bar{T}$ deformations.} 
Integrable structure underlying the KdV hierarchy in our AdS$_3$/CFT$_2$ setup suggests a possible connection to $T\bar{T}$-like deformations \cite{Zamolodchikov:2004ce,Smirnov:2016lqw,Cavaglia:2016oda,McGough:2016lol,Guica:2017lia,Bonelli:2018kik,Hartman:2018tkw,Cardy:2018sdv,Conti:2018tca,Apolo:2019yfj,LeFloch:2019rut,Guica:2020uhm,Leoni:2020rof,He:2020udl,Klaiber:2022ogf,Cardenas:2021sun,Apolo:2023aho,Babaei-Aghbolagh:2025lko,Aramini:2022wbn,Babaei-Aghbolagh:2024hti,Brizio:2024doe,Sakamoto:2025hwi}. In our construction, the functions $\mathcal{L}_\pm(t,x)$ encode the boundary stress tensor components via the Brown-Henneaux correspondence, and their dynamics is governed by the KdV hierarchy, an integrable system with an infinite set of commuting conserved charges. 

A natural question for further investigation is whether a similar integrability-based approach could be used to systematically derive $T\bar{T}$-like deformations of the boundary CFT. In particular, one may explore whether classical integrable flows, such as KdV or its forced/generalized variants, can provide a guiding principle for constructing irrelevant deformations that preserve certain conserved quantities. Reflectionless soliton solutions in the KdV context suggest that specific coherent excitations of the stress tensor can propagate without dispersion, and analogous structures may appear in integrable deformations of CFTs.

From the bulk perspective, these integrable flows correspond to trajectories in the space of allowed boundary conditions for AdS$_3$ metrics. Studying these trajectories in connection with $T\bar{T}$-type operators may shed light on how integrability constrains irrelevant deformations in holographic theories. We leave a detailed exploration of this connection for future work, as it may provide a systematic framework for relating classical soliton dynamics to integrable deformations in CFTs.

\paragraph{Extension to flat holography.}
A parallel question arises in the context of asymptotically flat gravity in three dimensions (see \cite{Grumiller:2017sjh} for the most general boundary condition), where the boundary dynamics is governed by the BMS$_3$ symmetry algebra and its associated Carrollian geometry. In this setting, the role of the functions $\mathcal{L}_\pm$ is replaced by the supertranslation and superrotation data that determine the null boundary stress tensor and current \cite{Adami:2024rkr}. The emergence of integrable structures in flat holography---including BMS$_3$ versions of the KdV and modified KdV hierarchies---has been observed in recent works, suggesting that similar integrability principles may underlie the dynamics of flat-space boundaries. 

It would be interesting to explore whether deformations analogous to $T\bar{T}$ (for instance \cite{He:2024yzx}), but adapted to the Carrollian or BMS-invariant setup, can be obtained from an integrability viewpoint similar to the one discussed here. Such deformations could correspond to consistent irrelevant flows preserving a subset of the infinite BMS$_3$ charges, or equivalently, to trajectories in the space of Carrollian boundary conditions of 3D flat gravity. Establishing this connection could provide a unified understanding of integrable deformations across AdS$_3$/CFT$_2$ and Flat/BFT$_2$ holography, and offer a bridge between solitonic dynamics, conserved charges, and irrelevant deformations in gravitational systems.

%%%%%%%%%%%%%%%%%%%%%%%%%%%%%%%%%%%%%%%%%%%%%%%%%%%%%%%%
\paragraph{Non-integrability.}
%%%%%%%%%%%%%%%%%%%%%%%%%%%%%%%%%%%%%%%%%%%%%%%%%%%%%%%%
The analysis presented in this work highlights a fundamental distinction between integrable and non-integrable driven systems, exemplified by the behavior of the KdV equation under external forcing. While generic driving terms destroy the Lax pair structure and the associated infinite hierarchy of conserved quantities, \emph{self-consistent sources} preserve integrability by coupling the source to the system’s own spectral data. This provides a concrete theoretical framework to probe and quantify weak integrability breaking in a controlled setting.

Within the AdS$_3$/CFT$_2$ correspondence, boundary dynamics governed by integrable structures such as the KdV hierarchy play a central role. An intriguing open question is how to interpret, on the gravitational side, a boundary evolution that transitions from an integrable, self-consistently driven regime to a non-integrable externally forced one. In particular, understanding how the loss of boundary conservation laws manifests in the bulk geometry could shed light on the relation between integrability breaking, chaotic dynamics, and horizon formation in asymptotically AdS$_3$ spacetimes.

%%%%%%%%%%%%%%%%%%%%%%%%%%%%%%%%%%%%%%%%%%%%%%%%%%%%%%%%%%%%%
\paragraph{Lax formalism and Pfaffian integrability.}
%%%%%%%%%%%%%%%%%%%%%%%%%%%%%%%%%%%%%%%%%%%%%%%%%%%%%%%%%%%%%
Our analysis has employed the Lax pair formalism as the operational criterion for integrability, enabling the systematic construction of conserved quantities and exact solutions. While this approach provides a powerful algebraic framework, it also lends itself to a broader geometric interpretation. In particular, integrability may equivalently be viewed as the complete integrability of a Pfaffian system \cite{freeman1984fully,kozlov1983integrability,Arnold:1989who,prigogine1995irreversibility,dubrovin2008hamiltonian,strachan2017frobenius} (see \cite{Ruzziconi:2020wrb} for gravity in three dimensions example), in the Frobenius sense, where the existence of sufficient conserved quantities corresponds to a foliation of the phase space by Lagrangian submanifolds. 

In this light, the zero-curvature condition underlying the Lax pair can be regarded as a specific realization of a Pfaffian integrability condition. The Lax connection, as a Lie-algebra–valued one-form, naturally defines such a system, while the spectral parameter may be interpreted as labeling a family of integrable Pfaffian structures. Clarifying this correspondence in the setting of three-dimensional gravity could provide a unified geometric perspective on integrability, bridging the algebraic Lax formulation with the symplectic and differential-geometric language of the phase space. This direction appears particularly promising for extending integrability concepts to other topological or higher-dimensional gravitational theories.

%%%%%%%%%%%%%%%%%%%%%%%%%%%%%%%%%%%%%%%%%%%%%%%%%%%%%%%%%%%%%
\paragraph{From classical to quantum Integrability.}
%%%%%%%%%%%%%%%%%%%%%%%%%%%%%%%%%%%%%%%%%%%%%%%%%%%%%%%%%%%%%
A natural next step is to explore how this classical structure can guide the quantization of the theory. The presence of an infinite hierarchy of conserved charges and the associated Lax connection indicates that the on-shell phase space possesses a well-structured symplectic geometry. Constructing an on-shell symplectic form explicitly remains a key open problem. In integrable systems, the Poisson algebra of monodromy matrices often exhibits a classical $r$-matrix structure, suggesting that the symplectic form might be expressible directly in terms of the underlying $r$-matrix, thereby linking the Lax representation to the canonical phase space geometry. 

With such a symplectic structure in hand, one could pursue geometric quantization. The integrable framework may naturally supply a suitable polarization—potentially arising from the complex structure on the spectral curve—allowing for the definition of physical wavefunctions as polarized sections of a prequantum line bundle. This program would provide a concrete realization of how the classical algebraic structures encoded in the Lax pair translate into quantum operators, such as quantized Wilson loops built from Lax monodromies. A comparison with alternative quantization approaches, including the Chern--Simons formulation and holographic constructions, could then offer valuable insights into the quantum structure of three-dimensional gravity.

%%%%%%%%%%%%%%%%%%%%%%%%%%%%%%%%%%%%%%%%%%%%%%%%%%%%%%%%%%%%%%%%%%%%%%%%%
\paragraph{Connection to higher-dimensional fluids.}
%%%%%%%%%%%%%%%%%%%%%%%%%%%%%%%%%%%%%%%%%%%%%%%%%%%%%%%%%%%%%%%%%%%%%%%%%
Our analysis of integrable structures in three-dimensional gravity via the Lax pair naturally suggests connections with recent developments in higher-dimensional fluid dynamics. In particular, \cite{Bustamante:2025wmi} shows that the dynamics of a classical, dissipationless compressible fluid in $(3+1)$-dimensions can be derived from a self-interacting Abelian Chern--Simons theory in $(4+1)$-dimensions, with integrability-like features such as helicity and entropy conservation emerging from the topological structure of the action. A compelling open question is whether the integrability of this higher-dimensional compressible fluid, characterized by constraints on the potential vorticity and new helicity invariants, can be formulated in terms of a Lax pair or another algebraic structure, revealing an underlying infinite-dimensional symmetry. Investigating the interplay between such topological field theory constructions and conventional notions of integrability in lower-dimensional systems represents a promising direction for future research.

%%%%%%%%%%%%%%%%%%%%%%%%%%%%%%%%%%%%%%%%%%%%%%%%%%%%%%%%%%%%%%%
\section*{Acknowledgments}
We would like to thank Ivan Sechin for fruitful discussions and Anatoly Dymarsky, Hern\'an A. Gonz\'alez and Daniel Grumiller for comments on the draft. We are also grateful to Babak Haghighat and Shing-Tung Yau for organizing the conference \emph{``Peaks, Depths, and Currents: Mathematics in Asia''}, where this work was initiated. We are grateful to the anonymous referee(s) for their insightful and constructive comments, which have significantly improved the clarity and quality of this work. H.A.\ additionally thanks M.~M.~Sheikh-Jabbari for long-term collaboration and insightful discussions on related topics. The work of HA is supported
by Beijing Natural Science Foundation under Grant No IS23018.

%%%%%%%%%%%%%%%%%%%%%%%%%%%%%%%%%%%%%%%%%%%%%%%%%%%%%%%%%%%%%%%%%%%

%%%%%%%%%%%%%%%%%%%%%%%%%%%%%%%%%%%%%%%%%%%
\bibliographystyle{fullsort.bst}
\bibliography{reference}

%%%%%%%%%%%%%%%%%%%%%%%%%%%%%%%%%%%%%%%%%%%

\end{document}